\documentclass[fleqn,usenatbib]{mnras}
\usepackage{newtxtext,newtxmath}
\usepackage[T1]{fontenc}
\DeclareRobustCommand{\VAN}[3]{#2}
\let\VANthebibliography\thebibliography
\def\thebibliography{\DeclareRobustCommand{\VAN}[3]{##3}\VANthebibliography}
\usepackage{threeparttable}
\usepackage{listings}
\usepackage[normalem]{ulem}
\usepackage{listings}
\usepackage{placeins}
\usepackage{graphicx}
\usepackage{color}	
\usepackage{amsmath}	
\usepackage{siunitx} 
\usepackage{multirow}
\usepackage{upgreek}
\definecolor{dkgreen}{rgb}{0,0.6,0}
\definecolor{gray}{rgb}{0.5,0.5,0.5}
\definecolor{mauve}{rgb}{0.58,0,0.82}
\usepackage{listings}
\lstdefinestyle{sqlstyle}{
  language=SQL,
  basicstyle=\ttfamily,
  keywordstyle=\color{blue},
  commentstyle=\color{green},
  stringstyle=\color{red},
  morekeywords={SELECT, FROM, WHERE, JOIN, ON}
}
\title[All-sky Catalogue of BeT Stars]{An all-sky catalogue of Be transition stars from WISE and Gaia}

\author[Qi Jia et al.]{
Qi Jia,$^{1}$
Mingjie Jian$^{2}$\thanks{E-mail: jian-mingjie@outlook.com},
Gang Li$^{3}$,
Qian Cui$^{4}$,
Chengyuan Li$^{1}$
\\
$^{1}$School of Physics and Astronomy, Sun Yat-sen University, Zhuhai 519082, China\\
$^{2}$Institute of Astronomy, University of Cambridge, Madingley Road, Cambridge CB3 0HA, UK\\
$^{3}$Centre for Astrophysics, University of Southern Queensland, Toowoomba, QLD 4350, Australia\\$^{4}$School of Physics and Astronomy, China West Normal University, Nanchong 637009, China\\}

\date{Accepted XXX. Received YYY; in original form ZZZ}

\pubyear{\the\year{}}

\begin{document}
\label{firstpage}
\pagerange{\pageref{firstpage}--\pageref{lastpage}}
\maketitle

\begin{abstract}
We present an all-sky catalogue of classical Be stars undergoing transitions between quiescent B-type and active Be-star phases, based on Gaia DR3 astrometry and multi-epoch mid-infrared photometry from WISE and NEOWISE-R. 
From a Gaia-selected parent sample of 4.2 million candidate OBA stars, we identify Be transition stars through correlated variability in $W_1$ magnitude and $W_1-W_2$ colour, which traces circumstellar disk formation and dissipation.
The final catalogue contains 3,300 Be transition stars, including 2,743 newly identified objects, and provides a large, homogeneous sample for studying Be outburst activity across the sky. 
The median durations of the measured outburst and decay phases are 459 and 478 days, respectively, consistent with previous WISE-based measurements and compatible with the long disk-evolution timescales expected in the viscous decretion disk framework.
As an initial application, we use TESS photometry for 3,017 targets to identify 63 eclipsing binary systems hosting Be transition stars, including 33 Algol-type, 17 Beta Lyrae-type, and 13 W Ursae Majoris-type systems.
Among the multi-epoch LAMOST spectra, two sources show candidate radial-velocity shifts exceeding 20 km s $^{-1}$; their line-profile behaviour suggests that RV variability in active Be transition stars may be influenced by outburst-related line-profile changes rather than being an unambiguous indicator of binary motion.
This catalogue provides a basis for future statistical studies of Be outbursts, disk evolution, and their connection with binarity and spectroscopic variability.
\end{abstract}

\begin{keywords}
stars: emission-line, Be -- stars: circumstellar matter -- stars: rotation -- catalogues -- surveys
\end{keywords}

\section{Introduction}

Classical Be stars are non-supergiant B-type stars that have exhibited Balmer line emission in their spectra \citep{1987pbes.coll....3C,2013A&ARv..21...69R}. 
This emission originates from a circumstellar environment of ionised gas \citep{1996A&A...308..170H,1997ApJ...479..477Q}, widely understood to be organised in a Keplerian rotating disk \citep{2007A&A...464...59M,2012MNRAS.423L..11W}. 
The prevailing viscous decretion disk (VDD) model explains the structure and evolution of these disks, which are formed and replenished by material ejected from the rapidly rotating central star during episodic outbursts, a process collectively known as the Be phenomenon \citep{1991MNRAS.250..432L,2001PASJ...53..119O,2008MNRAS.386.1922J,2011IAUS..272..325C,2012ApJ...756..156H}.

A critical phase for understanding this phenomenon is the transition between a quiescent B-type state and an active Be star state with a prominent disk \citep{2003PASP..115.1153P,2018AJ....155...53L}. 
These outbursts, characterised by enhanced mass transfer from the star to the disk, manifest as correlated photometric and spectroscopic variations \citep{2012ApJ...756..156H}. 
Because such transitions may be episodic and can occur on timescales of months to years, large and systematically selected samples are needed to characterise the timescales and diversity of disk formation and dissipation \citep{2024A&A...682A..59J}.
In this work, we refer to such objects as Be transition stars (hereafter BeT stars).

Previous studies of disk variability in Be stars have largely relied on known Be star catalogues, heterogeneous spectroscopic databases, or emission-line surveys \citep{1982IAUS...98..261J,2011AJ....142..149N,2018AJ....155...53L,2015MNRAS.446..274R}. 
Although these samples are valuable for detailed follow-up studies, they are affected by magnitude limits, uneven sky coverage, and survey-specific selection functions. 
Such selection effects make it difficult to assess the sky distribution, Galactic spatial distribution, and statistical properties of Be phase transitions in a uniform way.
A systematically selected all-sky sample is therefore needed to provide a more uniform basis for identifying Be transition stars and comparing their variability properties across different Galactic environments.

Mid-infrared observations provide a powerful complement to optical studies of Be star variability. 
This wavelength range is also particularly sensitive to emission from the circumstellar disk.
During disk formation and dissipation, changes in the disk contribution can therefore produce larger photometric variations in the mid-infrared than in the optical, because the disk contributes a larger fraction of the total flux at infrared wavelengths \citep{2012ApJ...756..156H,2018AJ....155...50G,2024A&A...682A..59J}. 
Compared with optical wavelengths, the mid-infrared is less affected by interstellar extinction, improving access to Be stars in the Galactic plane. 
The \textit{Wide-field Infrared Survey Explorer} (WISE; \citealt{2010AJ....140.1868W}) and its NEOWISE Reactivation phase (NEOWISE-R, \citealt{2014ApJ...792...30M}) provide an all-sky, multi-epoch mid-infrared dataset with a baseline exceeding 13 years, making them well suited for detecting the long-term brightness and colour variations characteristic of Be outbursts. 
Gaia DR3 provides the all-sky astrometry and photometry, together with uniformly processed stellar-parameter estimates, needed to define a large parent sample of candidate OBA-type stars \citep{2016A&A...595A...1G,2023A&A...674A...1G}.
Combining this Gaia-selected parent sample with long-baseline WISE/NEOWISE photometry then enables a systematic search for BeT stars across the sky.

In this paper, we present an all-sky catalogue of BeT stars selected from Gaia and WISE/NEOWISE data. 
Following the mid-infrared variability framework established by \citet{2024A&A...682A..59J}, we identify BeT candidates from long-term WISE light curves of a Gaia-selected OBA parent sample using correlated changes in $W_1$ magnitude and $W_1-W_2$ colour. 
We further perform ancillary characterisation of the resulting BeT sample using TESS and LAMOST data. 
High cadence \textit{Transiting Exoplanet Survey Satellite} \footnote{\url{https://tess.mit.edu}} (TESS; \citealt{2014SPIE.9143E..20R,2015JATIS...1a4003R}) photometry is used to search for photometric binary signatures. 
Multi-epoch spectra from the Large Sky Area Multi-Object Fiber Spectroscopic Telescope \citep[LAMOST;][]{2012RAA....12.1197C} are used to examine H$\alpha$ emission and candidate radial-velocity (RV) variability.

\section{Databases and sample selection}

This work uses four types of data, each serving a distinct role in our analysis. 
Gaia DR3 provides the all-sky astrometry, photometry, and stellar-parameter estimates used to construct the OBA-type parent sample.
WISE and NEOWISE multi-epoch photometry provide the long-baseline mid-infrared light curves used to identify disk-related variability.
TESS photometry and LAMOST spectra are used for ancillary characterisation, including searches for photometric binary signatures, H$\alpha$ emission, and candidate RV variability.
In this section, we describe the relevant survey data and the procedures for data retrieval, quality control, and sample construction.

\subsection{Gaia DR3 Parent Sample}

The \emph{Gaia} mission \citep{2016A&A...595A...1G} is a cornerstone ESA astrometric mission designed to map the Milky Way in three dimensions with unprecedented precision. 
Launched in 2013, \emph{Gaia} repeatedly scanned the entire sky until its science observations ended in January 2025, building an astrometric and photometric legacy for nearly two billion sources.
It provides absolute parallaxes, proper motions, and multi-band photometry in the $G$ (330--1050 nm), $G_{\mathrm{BP}}$ (330--680 nm), and $G_{\mathrm{RP}}$ (630--1050 nm) bands \citep{2018A&A...616A...1G,2018A&A...616A...4E}. 
For this work, we use Gaia DR3 stellar-parameter estimates \citep{2023A&A...674A...1G,2023A&A...674A..26C}. 
The effective temperatures, surface gravities, metallicities, and spectral types used here are derived from the GSP-Phot and ESP-HS pipelines \citep{2023A&A...674A..27A,2023A&A...674A..28F}.

Gaia DR3 provides a well-defined foundation for constructing the OBA parent sample required in this work. 
Ground-based spectroscopic surveys are often constrained by limited magnitude ranges and exhibit non-uniform sky coverage. 
In contrast, Gaia provides all-sky astrometry and photometry down to $G \sim 21$~mag, together with uniformly processed stellar-parameter estimates.

These attributes enable the construction of a large and consistently selected parent sample of candidate OBA-type stars across the sky. 
Accordingly, Gaia DR3 serves as the basis for the parent sample used in our subsequent mid-infrared variability analysis.

We construct the parent sample of candidate OBA stars from Gaia DR3 astrometric, photometric, and stellar-parameter data, using the ADQL query provided in Appendix \ref{adql}. 
We adapt the Gaia DR3 hot star selection strategy introduced by \citet{2023A&A...674A..37G} to select early type stars.
Stars are retained if they satisfy at least one of the following conditions: (1) GSP-Phot effective temperature $T_{\mathrm{eff}} > 7000\ \mathrm{K}$ and ESP-HS spectral type in ('O', 'B', 'A'); (2) ESP-HS effective temperature $T_{\mathrm{eff}} > 10000\ \mathrm{K}$ and GSP-Phot effective temperature $> 8000\ \mathrm{K}$; or (3) ESP-HS effective temperature in the range $10000\ \mathrm{K} < T_{\mathrm{eff}} < 50000\ \mathrm{K}$ when GSP-Phot parameters are unavailable.

To exclude nearby late type contaminants, we apply an absolute magnitude and colour cut. 
The condition is expressed as:

\[
\left(\frac{\varpi}{100}\right)^2 < 10^{2.0 - G + 1.8 \times (G_{\mathrm{BP}} - G_{\mathrm{RP}})},
\]

where $\varpi$ is the parallax in milliarcseconds, $G$ is the Gaia $G$ band magnitude, and $G_{\mathrm{BP}} - G_{\mathrm{RP}}$ is the Gaia BP-RP colour index. 
This cut removes nearby cool dwarfs with large parallaxes while retaining distant early type stars.
 
Applying the above criteria yields a parent sample of 4,241,582 candidate OBA stars with associated Gaia DR3 parameters, forming the basis for our subsequent mid-infrared variability analysis.

\subsection{WISE Time Series Photometry}\label{wisee}
We use multi-epoch mid-infrared photometry from the \textit{Wide-field Infrared Survey Explorer} (WISE; \citealt{2010AJ....140.1868W}) and its reactivation mission NEOWISE-R \citep{2014ApJ...792...30M} to trace the long-term variability of candidate OBA stars. 
The $W_1$ (3.4 $\mu$m) and $W_2$ (4.6 $\mu$m) bands provide a time baseline exceeding 13 years with a cadence of approximately six months, which is well suited for detecting the brightening and reddening associated with circumstellar disk formation, as well as the reverse trend during dissipation \citep{2012ApJ...756..156H,2018AJ....155...50G,2024A&A...682A..59J}.

For each source in the Gaia-selected parent sample, we retrieved and processed the single-exposure photometry from the ALLWISE and NEOWISE Reactivation databases following the procedures established by \citet{2024A&A...682A..59J}. 
This includes applying a $3^{\prime\prime}$ matching radius, flag-based quality cuts, saturation corrections, and binning the data by individual WISE/NEOWISE visits to produce robust light curves in $W_1$ magnitude and $W_1-W_2$ colour. 
After these quality cuts, we obtained multi-epoch light curves for 959,512 objects.

To identify Be transition (BeT) candidates, we adopted the methodology of \citet{2024A&A...682A..59J}, which employs a series of criteria based on correlated variability in the $W_1$ magnitude and $W_1-W_2$ colour. 
Specifically, we retained stars with mean $W_1$ magnitudes between 7 and 12.5 mag, required at least two epochs with reddened colours ($W_1-W_2 > -0.05 + 2\overline{\sigma_{W_1-W_2}}$), and demanded a peak-to-peak $W_1$ variation exceeding $3\overline{\sigma_{W_1}}$. 
We then computed the Pearson correlation coefficient ($r$) between the $W_1$ and $W_1-W_2$ time series. 
Following the empirical calibration of \citet{2024A&A...682A..59J}, we applied a threshold of $r < -0.70$ for the initial automated selection. 
This yielded 3,302 candidates, which were then subjected to a thorough visual inspection to exclude sources affected by instrumental artefacts or poorly sampled outbursts. 
Following this visual inspection, the final sample comprises 3,300 BeT candidates.

\subsection{Ancillary Data: TESS and LAMOST}
\subsubsection{TESS Observation}
The TESS mission \citep{2014SPIE.9143E..20R,2015JATIS...1a4003R} operates in a high cadence staring mode. 
It collects optical photometry continuously for $\sim27$ days per sector with a cadence of 30 minutes, or 2 minutes for selected targets. 
This observing mode is fundamentally different from the sparse sampling strategy of WISE. 
TESS captures variability on timescales from minutes to days, making it sensitive to photometric binary signatures such as eclipses and ellipsoidal variations. 
Although Be stars may also show short timescale pulsational variability, in this work we use TESS photometry primarily to search for binary signatures among the BeT stars identified from WISE.

We used the \texttt{Lightkurve} package \footnote{\url{https://lightkurve.github.io/lightkurve/}} \citep{2018ascl.soft12013L,2020AAS...23540904B} to retrieve TESS light curves for the 3,300 BeT candidates from multiple data products, including SPOC/TESS-SPOC \citep{2020RNAAS...4..201C}, QLP \citep{2020RNAAS...4..204H}, CDIPS \citep{2019ApJS..245...13B}, TASOC \citep{2021AJ....162..170H,2021ApJS..257...53L}, PATHOS \citep{2019MNRAS.490.3806N}, GSFC-ELEANOR-LITE \citep{2022RNAAS...6..111P}, and TGLC \citep{2023AJ....165...71H}. 
The SPOC and QLP products dominate the sample. 
Although data quality varies across pipelines, it is sufficient for identifying binary signatures. 
In total, we obtained light curves for 3,017 targets, corresponding to 91.4 per cent of the 3,300 BeT stars.

Among the retrieved light curves, QLP products account for 69.5 per cent, TESS-SPOC for 22.4 per cent, GSFC-ELEANOR-LITE for 5.4 per cent, TASOC for 1.5 per cent, CDIPS for 0.9 per cent, and SPOC and TGLC for the remainder. 
We caution that the SPOC and TESS-SPOC light curves are the most reliable TESS data products, whereas the other pipeline products may contain residual systematics; these are nevertheless adequate for identifying the eclipses of binary systems. 
All binary candidates were visually inspected, and for targets observed by more than one pipeline we verified that the eclipse signals are consistent between the different data products. 
The pipeline adopted for each target is recorded in the Author column of Table \ref{ob}.

Based on the available TESS light curves, we searched for eclipsing binary systems among the BeT stars. 
We converted the normalised fluxes into relative TESS magnitudes for subsequent analysis using

\( m = -2.5 \times \log_{10}(\text{flux}) + m_\mathrm{shift} \),

where $m_\mathrm{shift}$ is an arbitrary magnitude offset adopted only to facilitate visual comparison with WISE photometry. 

We then analysed periodic signals using the Lomb–Scargle method \citep{1976Ap&SS..39..447L,1982ApJ...263..835S}. 
For each target, we recorded the frequency corresponding to the highest peak in the periodogram and used its reciprocal as an initial candidate period. 
However, the highest peak in the periodogram does not always correspond to the true orbital period, especially for eclipsing binaries where aliases at half or twice the true period frequently occur. 
To mitigate this issue, we visually inspected every folded light curve. 
When the folded light curve exhibited a double dip or asymmetric eclipse pattern inconsistent with the initial period, we manually tested integer multiples, such as 2 $\times$ or 0.5 $\times$ of the candidate period. 
The final adopted orbital period was the one that yielded the most coherent and physically plausible folded light curve morphology. 

Using the adopted period, we folded the light curve and fitted it with a sixth order Fourier series:

\begin{equation}
    m (\phi)  = a_0 + \sum_{i=1}^{6} a_i \cos(2\pi i \phi ) + \sum_{i=1}^{6} b_i \sin(2\pi i \phi )
    \label{eq: fourier}
\end{equation}

Here $m (\phi)$ is the differential magnitude at orbital phase $\phi$; and \(a_0\) is the mean magnitude level over the orbital cycle, \(a_i\) and \(b_i\) are the Fourier coefficients of the \(i\)-th harmonic component.
The Fourier model provides a compact description of the folded light curve morphology, including eclipse depth, eclipse width, and ellipsoidal variations \citep{2020ApJS..249...18C}.
We then visually inspected the light curve, periodogram, and folded light curve for each candidate signal, ultimately identifying 63 objects with convincing eclipsing or EB-like light-curve morphology.

\subsubsection{LAMOST Observation}

LAMOST \citep{2012RAA....12.1197C,2012RAA....12..723Z} is a quasi meridian reflecting Schmidt telescope with 4000 fibres distributed over its focal plane, enabling simultaneous multi object spectroscopy over a wide field of view. 
Its low resolution spectroscopic (LRS) survey provides spectra with a resolving power of $R\sim1800$ over the wavelength range 3700--9000 \AA. 
This wavelength coverage includes H$\alpha$, making LAMOST well suited for identifying the spectroscopic signatures of circumstellar disk activity in Be stars. 
In classical Be stars, the H$\alpha$ line at 6562.85 \AA\ appears in emission when a dense circumstellar disk is present, and its equivalent width traces disk activity and density \citep{1996A&A...308..170H}. 
Multi epoch spectroscopy is therefore useful for capturing the transient nature of Be outbursts, as a single spectrum only provides a snapshot of whether a disk exists at a given moment.

We cross-match our BeT sample with LAMOST Data Release 13 \footnote{\url{https://www.lamost.org/dr13/}} LRS to search for spectroscopic variability. 
This cross-match yields 486 unique targets with at least one LAMOST LRS spectrum. 
Among these, 154 sources have two or more epochs of spectroscopic observations, providing the multi-epoch sampling needed to examine temporal changes in H$\alpha$ emission and RVs. 
A recent LAMOST DR13 study by \cite{2026ApJS..283...59A} has already analysed a subset of classical Be stars and candidate Be star binaries using multi-epoch LAMOST spectra. 
Of our 154 multi-epoch sources, 67 overlap with that study; we therefore focus our new RV measurements on the remaining 87 sources.

For the remaining 87 multi-epoch LRS sources, we performed RV measurements. 
Given the low resolving power of LAMOST LRS ($R \sim 1800$), detailed line profile modelling is not feasible. 
We therefore adopted a pragmatic approach to estimate relative RV shifts between epochs.
We measured RVs from several relatively clean photospheric lines: H$\gamma$ (4340.47\,\AA), H$\beta$ (4861.35\,\AA), H$\alpha$ (6562.85\,\AA), H$\delta$ (4101.73\,\AA), and a set of He\,I lines (4026.19, 4387.92, 4921.93, 6678.15, 7065.25\,\AA). 
At this resolution, the line profiles are dominated by the instrumental line spread function.
We therefore adopted a Gaussian profile as a simple approximation for fitting the line cores and measuring centroid shifts. 
For each selected line, the Gaussian centroid was used to determine the RV, and the associated uncertainty was estimated with a Markov chain Monte Carlo (MCMC) sampler \citep{2013PASP..125..306F}. 
For H$\alpha$, we used the line for RV measurements only when it appeared as a clear absorption feature. 
H$\alpha$ profiles dominated by emission or complex disk-related structure were excluded, because their centroids may not trace the stellar RV.

We classified a star as a candidate RV variable if the difference in its measured RV between any two epochs exceeded 20 km s $^{-1}$.
This threshold follows the numerical criterion adopted by \cite{2026ApJS..283...59A} for LAMOST medium-resolution spectroscopy and is used here as a pragmatic selection threshold for candidate RV variability in the LRS data.

\section{Results}

\subsection{The all-sky catalogue of BeT stars}

\begin{figure*}
\includegraphics[width=\linewidth]{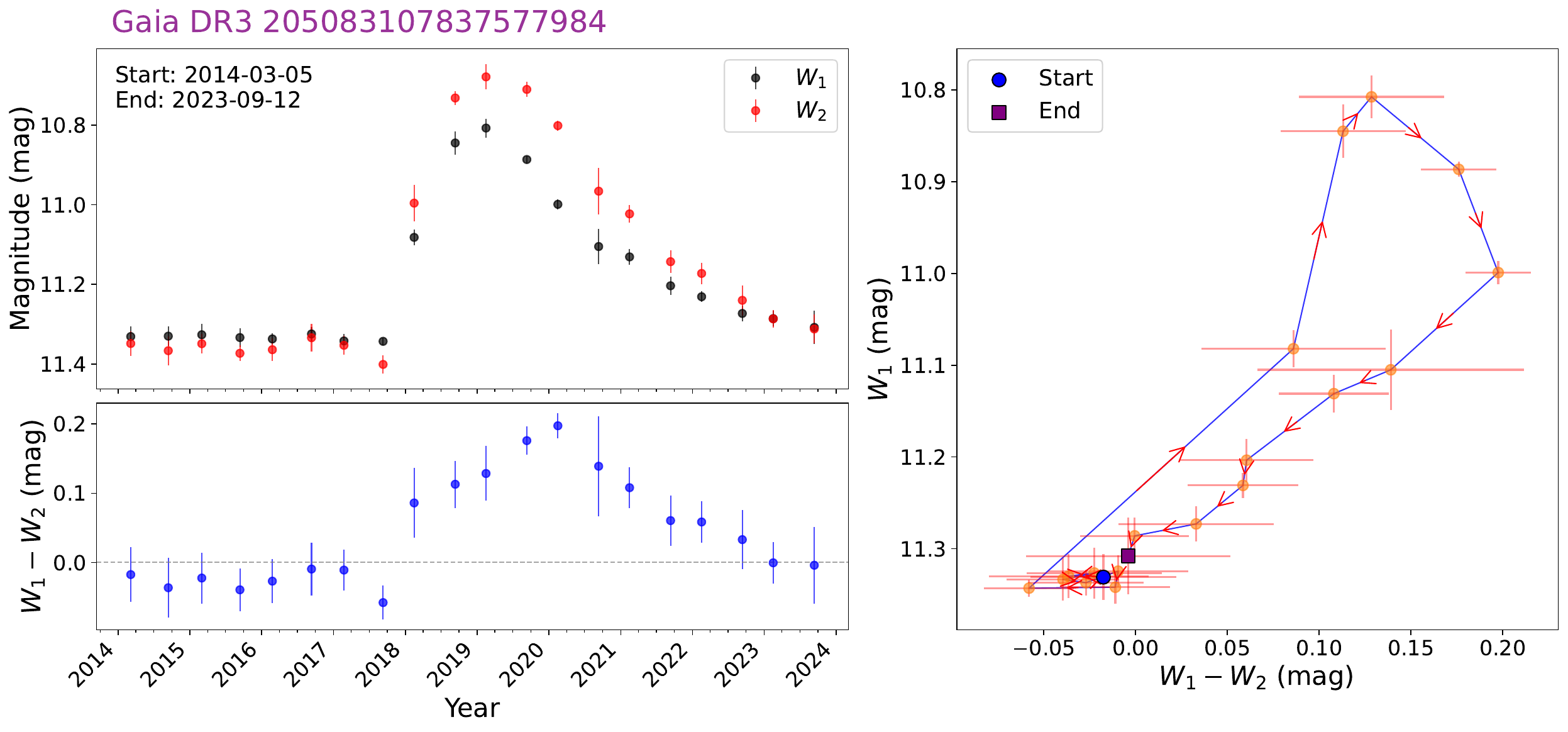}
\caption{WISE light curve of the BeT star Gaia DR3 205083107837577984. Left: WISE $W_1$ and $W_2$ light curves (top) and the temporal evolution of the $W_1 - W_2$ colour (bottom). Right: CMD constructed from WISE epoch photometry.} \label{lc1}
\end{figure*}

\begin{figure*}
	\includegraphics[width=\linewidth]{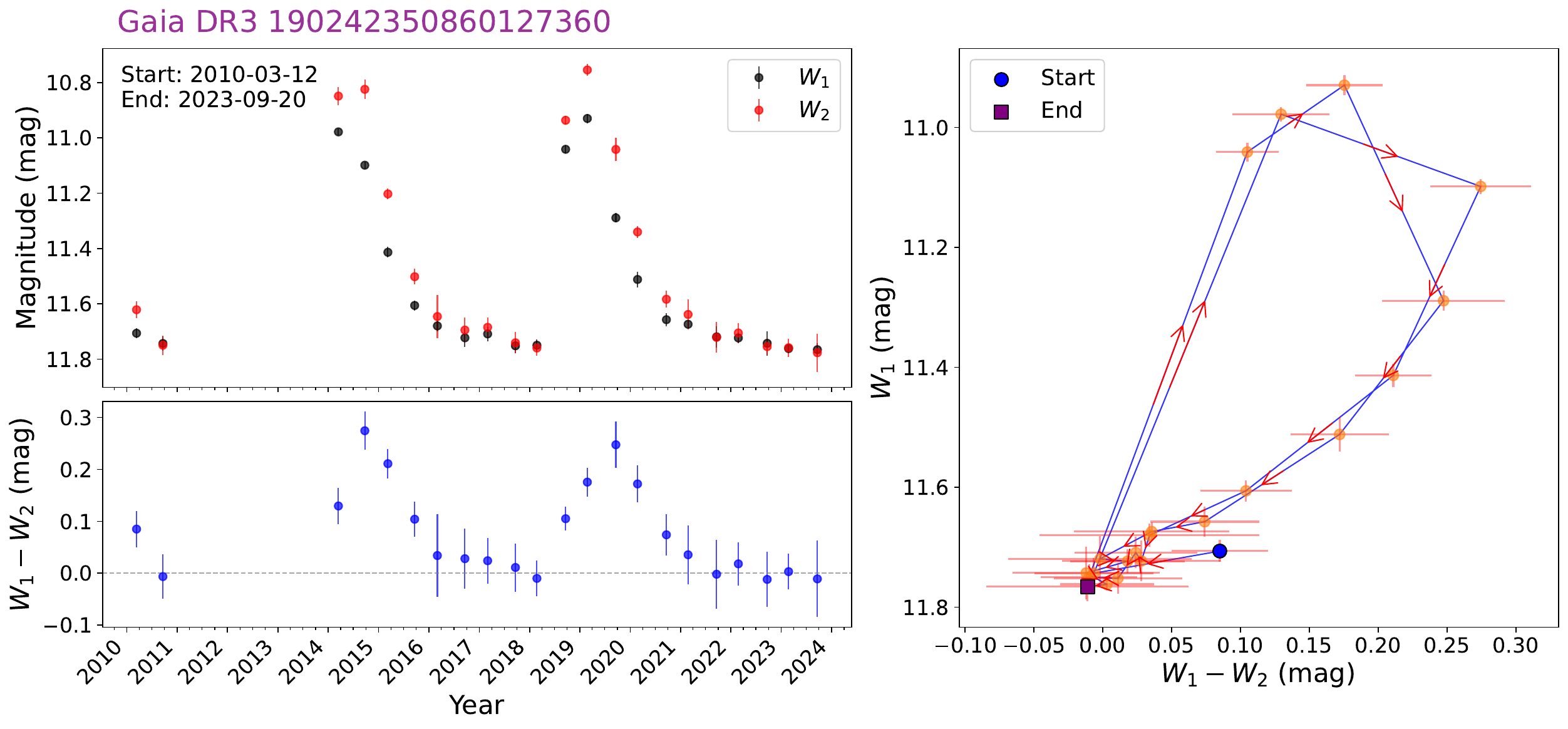}
    \caption{WISE light curve of the BeT star Gaia DR3 190242350860127360. Same as Fig. \ref{lc1}, but showing two complete outburst cycles.}
    \label{lc2}
\end{figure*}

\begin{figure*}
	\includegraphics[width=\linewidth]{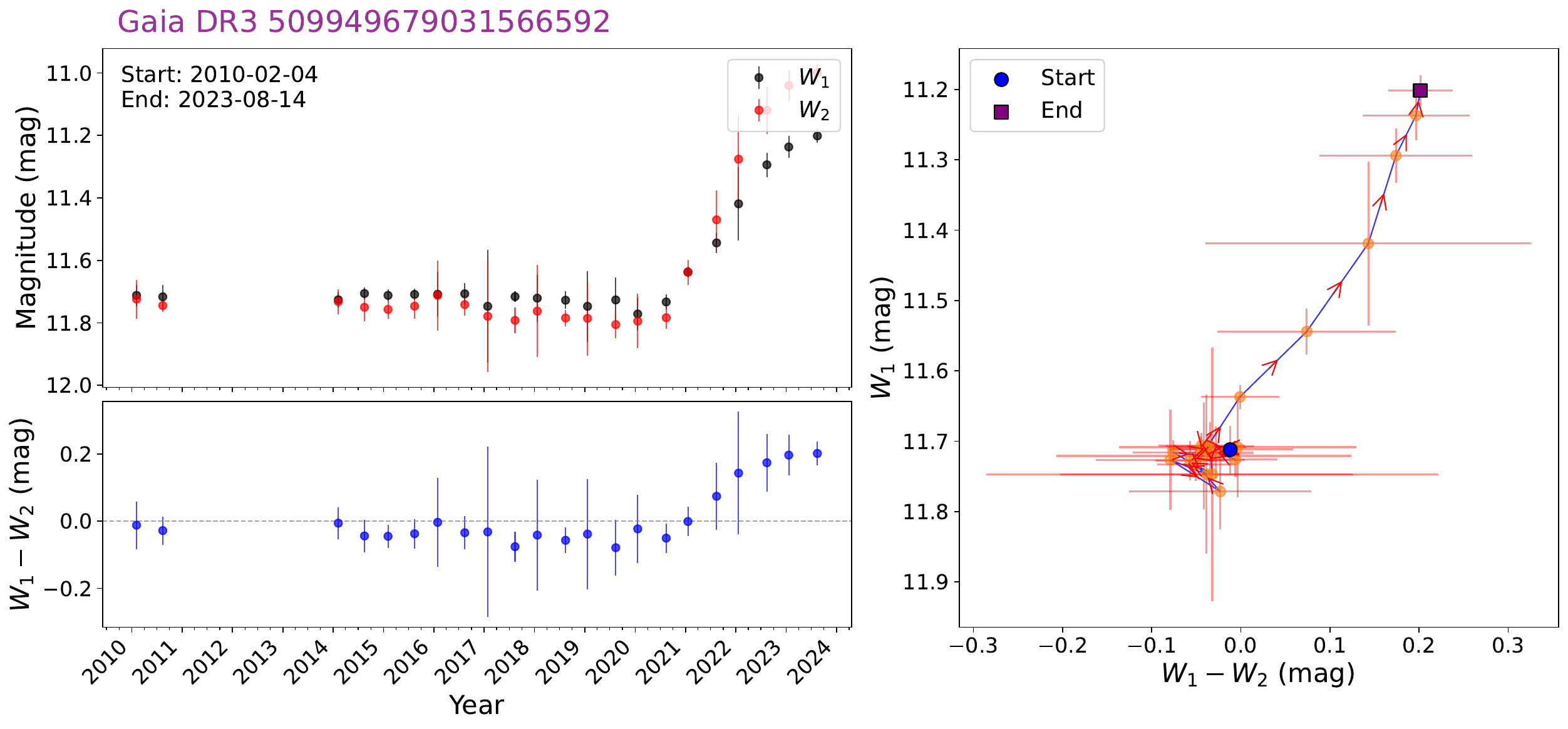}
    \caption{WISE light curve of the BeT star Gaia DR3 509949679031566592. Same as Fig. \ref{lc1}, but showing only the transition from the quiescent phase to the outburst phase.}
    \label{lc3}
\end{figure*}

\begin{figure*}
	\includegraphics[width=\linewidth]{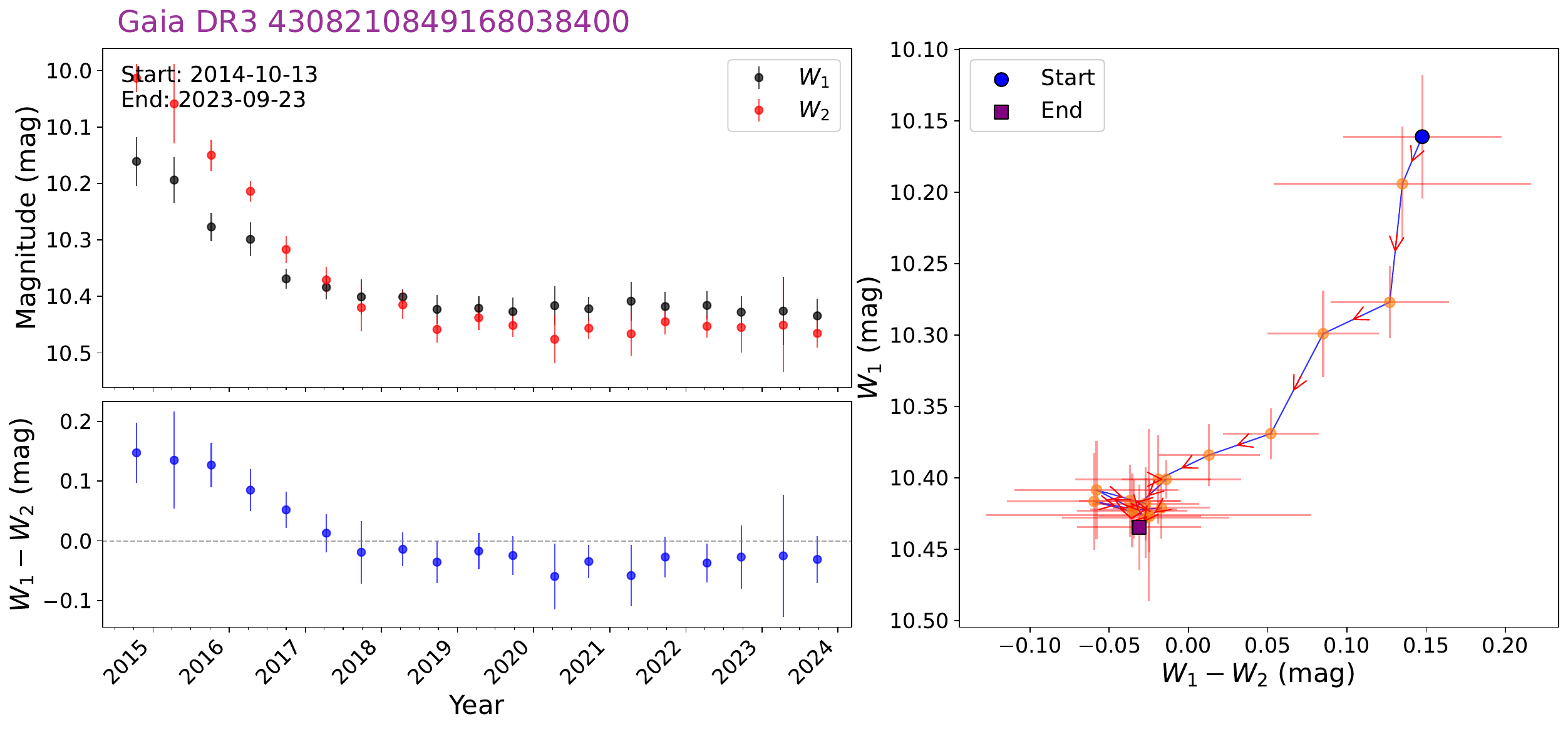}
    \caption{WISE light curve of the BeT star Gaia DR3 4308210849168038400. Same as Fig. \ref{lc1}, but showing only the transition from the outburst phase to the quiescent phase.}
    \label{lc4}
\end{figure*}

We identify a final sample of 3,300 BeT stars. 
Figure \ref{lc1} shows an example of a BeT star light curve and its corresponding CMD. 
Figure \ref{lc2} captures two complete outburst cycles from the same star, illustrating that Be stars can undergo recurrent disk formation and dissipation events. 
Figures \ref{lc3} and \ref{lc4} show examples in which only the rising or decaying branch of an individual outburst is captured.
These examples illustrate that some outbursts are only partially sampled by WISE/NEOWISE; even with a long time baseline, the full evolution of individual events can be missed because of the sparse cadence.
The light curves and corresponding figures for all candidates are publicly available via China-VO \footnote{\url{https://doi.org/10.12149/101344}}.

\begin{figure*}
	\includegraphics[width=\linewidth]{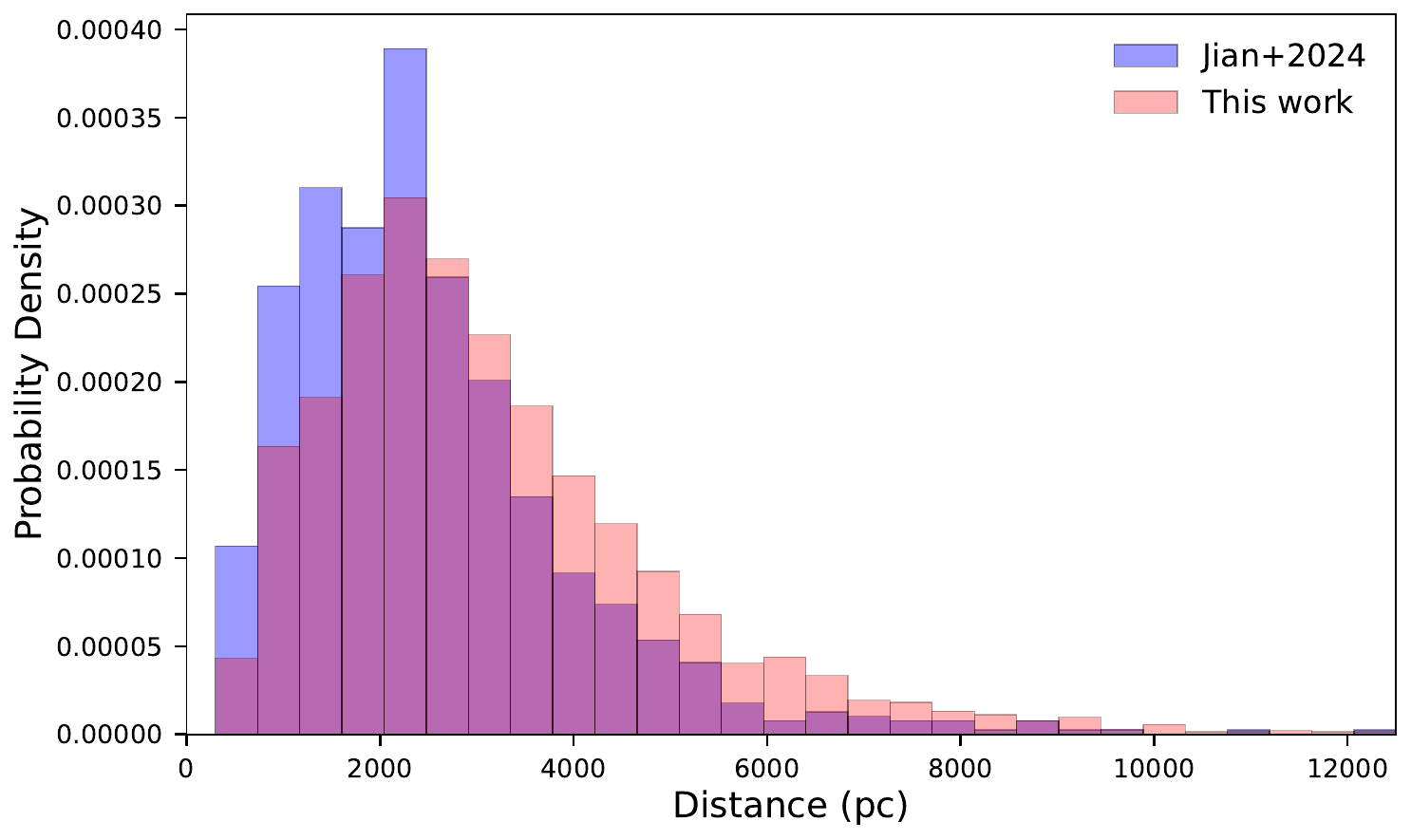}
    \caption{Normalised distance distributions of the BeT stars in this work (red) and of \protect\citet{2024A&A...682A..59J} (blue). Distances for both samples are the Gaia geometric distances of \protect\citet{2021AJ....161..147B}.}
    \label{dis}
\end{figure*}

Selected catalogue entries are shown in Table \ref{ob}, and Figure \ref{dis} compares the distance distributions of our sample and that of \cite{2024A&A...682A..59J}, showing that our sample extends to larger heliocentric distances.

We cross-matched our final sample with the catalogue of \cite{2024A&A...682A..59J} and recovered 557 common sources, while identifying 2,743 additional BeT stars.
Although our all-sky Gaia-selected parent sample might be expected to include the previously identified BeT stars, it does not recover all sources in the catalogue of \cite{2024A&A...682A..59J}. 
This is primarily because the two studies adopt different input star catalogues and selection functions.
\cite{2024A&A...682A..59J} based their search on spectroscopically confirmed OBA-type stars from multiple surveys, whereas our parent sample is selected from Gaia DR3 data. 
Spectroscopic input catalogues are often biased toward brighter targets and specific sky regions, while the Gaia-selected sample provides broader all-sky coverage, albeit with less direct spectroscopic validation.

\begin{table*}
\centering
\begin{threeparttable}
\caption{Sample entries from the all-sky BeT catalogue.\label{ob}}
{\footnotesize
\centering
\begin{tabular*}{\textwidth}{@{\extracolsep{\fill}}lccccccccccc}
\hline
DR3 Name & TIC & R.A.(J2000) & Dec.(J2000) & $({W_1-W_2})_{\mathrm{ptp}}$ & ${W_1}_{\mathrm{ptp}}$ & ... & $P$ & Type & Author\\
 & & (deg) & (deg) & (mag) & (mag) & & (days) & & \\
 \hline
Gaia DR3 187236011192027776 & 187632813   & 78.34391 & 37.11895 & 0.0715 & 0.1755 & ... & 1.8220 & 3 & QLP            \\
Gaia DR3 207611198605353344 & 98620667    & 79.78155 & 43.79550 & 0.0557 & 0.3024 & ... & 1.7333 & 3 & TESS-SPOC      \\
Gaia DR3 3105336847653580416 & 280217676 & 100.86442 & -3.93051 & 0.3300 & 0.8375 & ... & 0.7742 & 2 & QLP            \\
Gaia DR3 5890905816704254976 & 291429783 & 217.94150 & -58.88950 & 0.1371 & 0.3156 & ... & 0.2205 & 2 & TESS-SPOC     \\
Gaia DR3 5308882222753920512 & 440962587 & 144.98876 & -54.10765 & 0.1205 & 0.1665 & ... & 1.5508 & 2 & GSFC-ELEANOR-LITE \\
Gaia DR3 5258571418203139712 & 463591205 & 154.26300 & -58.14843 & 0.1848 & 0.3872 & ... & 6.2112 & 2 & QLP            \\
Gaia DR3 5715632187060272768 & 141010371 & 116.68716 & -20.66469 & 0.3120 & 1.0520 & ... & 0.7477 & 1 & QLP            \\  
Gaia DR3 4662147318644179584 & 30275518   & 73.96045 & -66.57582 & 0.1985 & 0.2175 & ... & 2.2110 & 1 & TGLC           \\
Gaia DR3 2059002275145707392 & 90978777   & 301.93539 & 35.35932 & 0.3250 & 0.2450 & ... & 1.4179 & 1 & QLP            \\
Gaia DR3 474417036635830784  & 392316838 & 55.01043 & 60.95242 & 0.0545 & 0.1070 & ... & 5.1802 & 1 & TESS-SPOC        \\        
Gaia DR3 5966504044325373440 & 339566181 & 253.47716 & -41.87084 & 0.2367 & 0.5890 & ... & 4.2329 & 1 & QLP            \\
Gaia DR3 1928824152839954432 & 129281903 & 342.52395 & 38.41033 & 0.1450 & 0.7565 & ... & -      & - &   -             \\
Gaia DR3 2178230876502184576 & 342333765 & 326.59076 & 57.13997 & 0.2015 & 0.3675 & ... & -      & - & QLP             \\
Gaia DR3 1972937624504810368 & 190696940 & 325.64552 & 43.26160 & 0.1939 & 0.2283 & ... & -      & - & TESS-SPOC       \\
Gaia DR3 4173269505072942336 & 167345277 & 274.43817 & -5.03711 & 0.0970 & 0.0450 & ... & -      & - & QLP             \\
Gaia DR3 5326406376492885760 & 293551253 & 139.32740 & -48.17581 & 0.1590 & 0.3545 & ... & -      & - & GSFC-ELEANOR-LITE \\
Gaia DR3 2172448338690394368 & 64382173   & 321.75162 & 51.68450 & 0.4255 & 1.1755 & ... & -      & - & QLP            \\
Gaia DR3 449281380488462080  & 316246047 & 51.82458 & 56.62153 & 0.1185 & 0.3025 & ... & -      & - &  -               \\
Gaia DR3 4309657840813094400 & 320454065 & 290.51759 & 11.29316 & 0.1680 & 0.3660 & ... & -      & - & TESS-SPOC       \\
Gaia DR3 5523570870308125952 & 141577846 & 129.89730 & -42.63185 & 0.2490 & 0.6250 & ... & -      & - & GSFC-ELEANOR-LITE \\
Gaia DR3 184503106320806528  & 143758066 & 80.81207 & 37.71490 & 0.0705 & 0.1835 & ... & -      & - & TESS-SPOC        \\
Gaia DR3 2008514797121206144 & 317156275 & 342.02687 & 59.70388 & 0.2250 & 0.5820 & ... & -      & - & CDIPS           \\
Gaia DR3 6053725278124148992 & 450563844 & 186.59797 & -63.56447 & 0.0415 & 0.0860 & ... & -      & - & GSFC-ELEANOR-LITE \\
Gaia DR3 4321606233664995200 & 69210724   & 291.89253 & 16.08997 & 0.2300 & 0.2505 & ... & -      & - &    -           \\
Gaia DR3 5619567615624757504 & 139561907 & 112.00776 & -21.91498 & 0.0775 & 0.2250 & ... & -      & - & QLP            \\
Gaia DR3 2018376316912394752 & 423543930 & 294.82281 & 22.08924 & 0.1230 & 0.2120 & ... & -      & - &    -            \\
...  & ...   & ...  & ...     & ...  & ...    & ... & ...   &...  & ...\\
Gaia DR3 139265796338853888  & 177824257 & 48.08020 & 36.99640 & 0.1366 & 0.1607 & ... & -      & - & TESS-SPOC        \\
\hline
\end{tabular*}
}
\begin{tablenotes}
	\item{\bf Notes.} DR3 Name: Gaia DR3 source identifier; TIC: TESS input catalogue identifier; R.A. and Dec.: source position in J2000 coordinates; $({W_1-W_2})_{\mathrm{ptp}}$ and ${W_1}_{\mathrm{ptp}}$: Peak to peak variations of $W_1 - W_2$ and $W_1$, respectively; $P$: adopted orbital period; Type: binary morphology type, with 1 = EA, 2 = EB, 3 = EW; Author: TESS data pipeline; "-" indicates unavailable values.\\ 
\end{tablenotes}
\end{threeparttable}
\end{table*}

\begin{figure*}
\includegraphics[width=\linewidth]{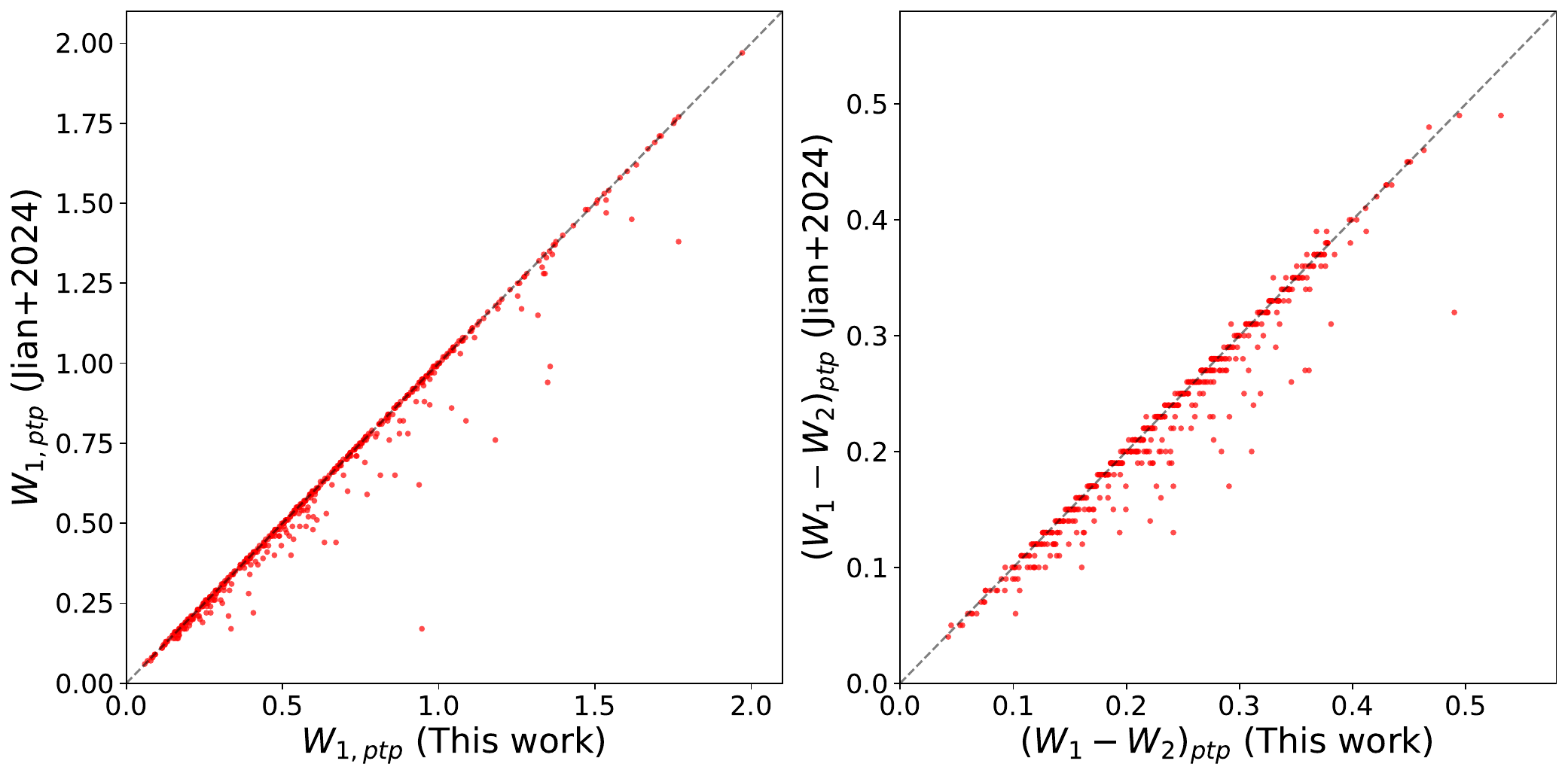}
    \caption{Comparison of $W_1$ and $W_1-W_2$ peak-to-peak amplitudes between this work and \protect\cite{2024A&A...682A..59J}.}
    \label{cmptp}
\end{figure*}

To assess the consistency between this work and \cite{2024A&A...682A..59J}, we compared the peak to peak variations in $W_1$ and $W_1 - W_2$ for the 557 common sources.
As shown in Figure \ref{cmptp}, most sources show broadly consistent amplitudes between the two studies.
However, the amplitudes measured in this work are often larger than those reported by \cite{2024A&A...682A..59J}.
This difference is likely due to the longer WISE/NEOWISE time baseline and the inclusion of additional epochs in this work, which allow us to sample a wider range of variability.

\begin{figure*}
	\includegraphics[width=\linewidth]{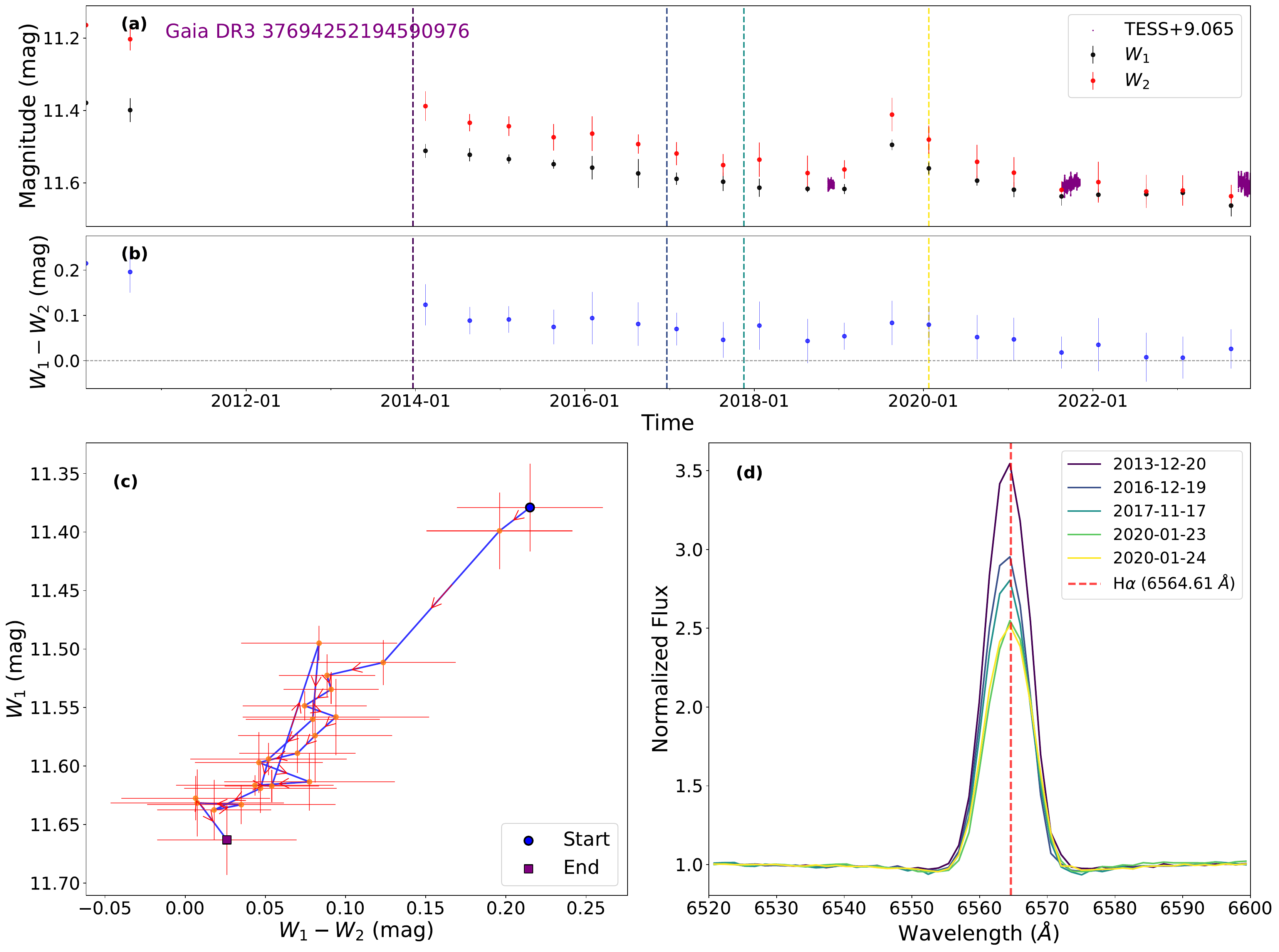}
    \caption{Combined WISE, TESS, and LAMOST view of Gaia DR3 37694252194590976 (TIC 435336653). (a) WISE and TESS light curves; vertical lines indicate LAMOST spectroscopic observation dates. (b) Temporal evolution of the WISE $W_1 - W_2$ colour. (c) CMD from WISE epoch photometry. (d) Time-varying H$\alpha$ line profiles from LAMOST spectra of the same target.}
    \label{combined}
\end{figure*}

As an illustrative multiwavelength example of the catalogue, we selected a target with frequent observations from both TESS and LAMOST, as shown in Figure \ref{combined}. 
The available LAMOST spectra show clear H$\alpha$ emission during epochs covered by the WISE light curve. 
Together with the WISE and TESS photometry, these data illustrate the connection between disk activity and line emission, although a quantitative analysis of H$\alpha$ equivalent width evolution is beyond the scope of this work.

\subsection{O/D Duration Statistics}
Following the methodology established by \citet{2024A&A...682A..59J}, we analysed the durations of all identified outburst (O) and decay (D) events in the BeT light curves. 
These phase durations provide useful constraints on the timescales of circumstellar disk formation and dissipation, which are key quantities in the VDD framework \citep{2012ApJ...756..156H,2013A&ARv..21...69R}.
We adopt the $W_1$ band as the primary band for measuring phase durations because its measurements have smaller photometric uncertainties and are less affected by thermal background emission than $W_2$ \citep{2014ApJ...792...30M}, and because this allows a direct comparison with the phase durations of \citet{2024A&A...682A..59J} measured in the same band.

To extract the phase durations, we fitted a smoothing B-spline to the WISE $W_1$-band light curve of each star. 
The fitted curve is expressed as

\begin{equation}
f(t) = \sum_{j} c_j B_{j,k}(t),
\end{equation}

where $B_{j,k}(t)$ are the B-spline basis functions of degree $k$ defined on a knot vector, and the coefficients $c_j$ are determined using the FITPACK routine \texttt{curfit} \citep{1993csfw.book.....D}. 
The smoothing parameter $s$ controls the trade-off between fidelity to the data and smoothness of the fitted curve by imposing the constraint

\begin{equation}
\sum_{i} w_i \left[ W_1(t_i) - f(t_i) \right]^2 \le s,
\end{equation}

where $w_i$ are the weights. Larger values of $s$ permit greater residuals and therefore produce smoother curves that suppress high-frequency noise, while smaller values of $s$ force a closer fit to the individual measurements. 
We adopted $s = 0.04$ throughout the analysis.

After smoothing, we identified consecutive monotonic segments in the $W_1$ light curve as outburst (brightening) and decay (fading) phases, respectively. 
Only phases with total $W_1$ changes exceeding 0.1 mag and durations longer than 180 days were retained. 
Phases within 90 days of the observational baseline endpoints were discarded to avoid incomplete sampling, and phases containing gaps longer than 3 years (1095 days) in the original data were excluded to ensure sufficient temporal coverage of the evolutionary sequence.

\begin{figure}
	\includegraphics[width=\linewidth]{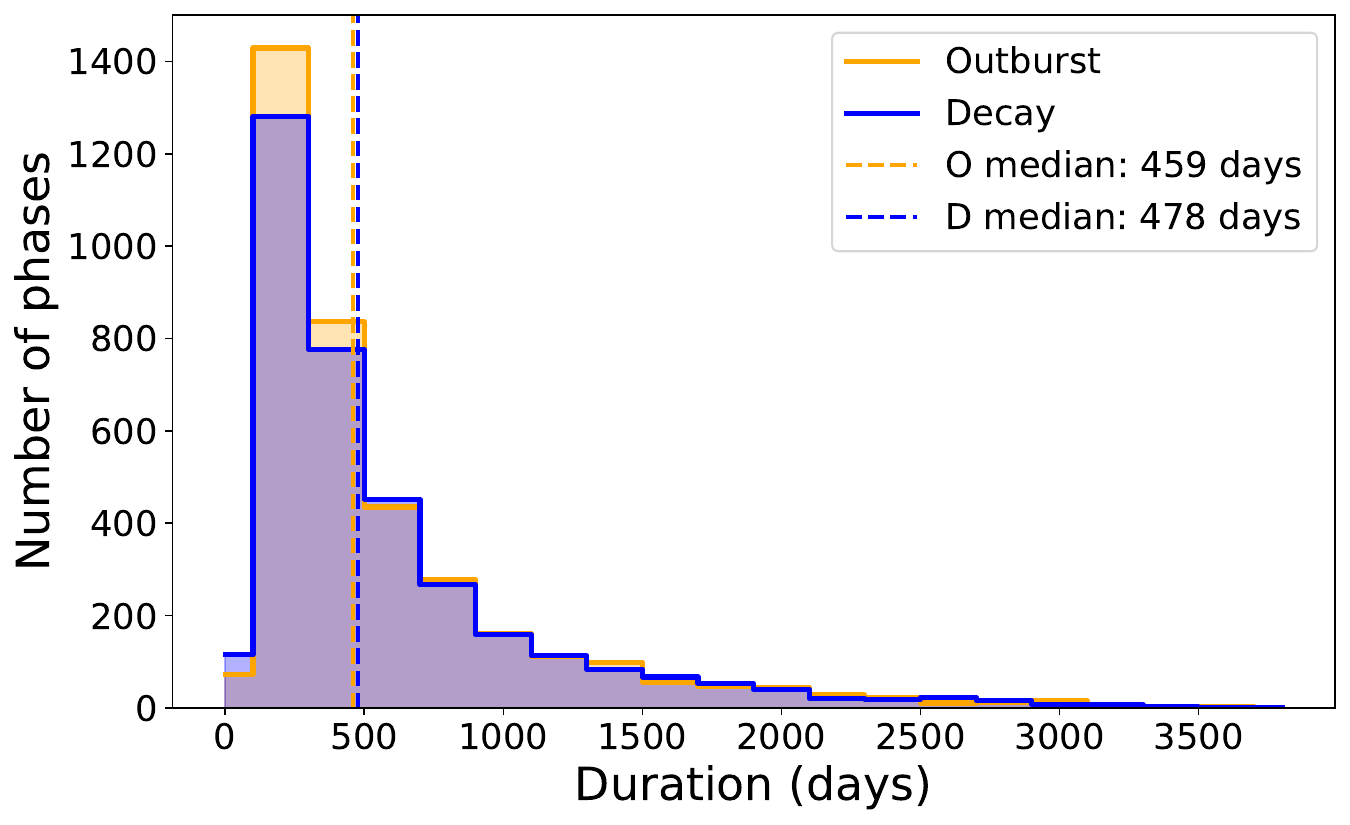}
    \caption{Distribution of all measured O and D phase durations. The blue and orange histograms represent the O and D events, respectively. The vertical lines mark the median durations of the measured events (O: 459 days; D: 478 days).}
    \label{od}
\end{figure}

Figure \ref{od} presents the duration distributions of all measured O and D events. 
Considering these O and D events independently, the median durations are 459 and 478 days, respectively.
These values are broadly consistent with the medians reported by \citet{2024A&A...682A..59J} (474 and 524 days, respectively), given the typical measurement uncertainty of approximately $\pm90$ days imposed by the $\sim6$ month sampling cadence of WISE/NEOWISE.
Both phases span a wide range from approximately 200 to 1800 days, with a prominent peak around 300--500 days. 
This broad distribution reflects the diversity of disk evolution timescales among Be stars, potentially related to differences in disk viscosity, mass injection history, and viewing inclination \citep{2012ApJ...756..156H,2018AJ....155...50G}.

To verify the robustness of adopting $W_1$ as the primary band, we repeated the phase identification and duration measurement using the $W_2$ light curves with the same spline settings. 
In total, $W_2$ yields about 21 per cent more detected phases than $W_1$ (8,666 versus 7,171). 
The larger photometric uncertainties in $W_2$ make the spline fits less stable for some sources and can introduce additional low-significance or poorly constrained O/D segments.
For the 3,544 O phases and 3,297 D phases that are well detected in both bands, the durations derived from the two bands are broadly consistent: the median durations measured in $W_1$ exceed those in $W_2$ by about 16 days for the O phases and 11 days for the D phases, corresponding to median $W_1/W_2$ ratios of 1.03 and 1.02. 
Thus, for phases reliably identified in both bands, the measured durations are insensitive to the adopted band, while $W_1$ provides more stable phase identification.

We also examined the long-duration tail of the distributions. 
For the 142 events (73 O and 69 D phases, associated with 96 stars) with durations exceeding 1000 days, we individually inspected the light curves and spline segmentations. 
Eleven events (7.7 per cent) are located at the boundaries of the observational baseline or span the $\sim$3-year gap between the WISE mission phases, and their durations are therefore uncertain; the remaining 131 events show clear, continuous brightening or fading trends with reasonable spline segmentation and no mis-segmentation caused by anomalous data points. 
We further compared our duration distributions with those of \citet{2024A&A...682A..59J} (2,204 O and 2,091 D phases) using two-sample Kolmogorov--Smirnov tests. 
The resulting $p$-values of 0.31 for the O phases and 0.27 for the D phases do not reject the hypothesis that the two samples are drawn from the same distribution, and the median durations of the two studies agree to within 50 days, well below the $\sim$180-day sampling interval. 
Together with the individual inspection of the long-duration events, this comparison indicates that the extended tail is not dominated by obvious mis-segmentation or anomalous data points, and is consistent with a population of genuinely long disk-evolution events.

The measured timescales are compatible with the long term disk evolution expected in the VDD framework, in which disk material redistributes through viscous diffusion \citep{2001PASJ...53..119O,2013A&ARv..21...69R,2012ApJ...756..156H}.
However, because phases shorter than 180 days were excluded and the WISE cadence limits the time resolution, these durations should be interpreted as a statistical characterisation of long term disk evolution rather than precise event by event measurements of disk formation and dissipation timescales.
The sparse sampling also leads to a second type of incompleteness: if a Be star remained in a stable disk or diskless state throughout the 13-year WISE baseline, or if the quiescent interval between successive outbursts exceeded the baseline, the star would not exhibit identifiable correlated brightness--colour variations and would be missed by our selection. 
We stress that not being identified as a BeT star does not imply that a star is not a classical Be star; this selection effect is quantified through the recovery rate of BeSS reference stars in Section \ref{purity}.

\subsection{BeT Stars in Eclipsing Binary Systems}

Based on the folded light curves and the adopted orbital periods, we classified each eclipsing binary system according to the conventional morphological types of eclipsing binaries \citep{2017ARep...61...80S}. 
Type 1 corresponds to EA (Algol-type) systems, which show well separated primary and secondary eclipses and relatively flat light curves outside eclipse. 
Type 2 corresponds to EB (Beta Lyrae-type) systems, which show continuously varying light curves, often associated with ellipsoidal distortion and eclipses of different depths. 
Type 3 corresponds to EW (W Ursae Majoris-type) contact binaries, which show nearly sinusoidal light curves with minima of similar depth.

\begin{figure*}
\includegraphics[width=\linewidth]{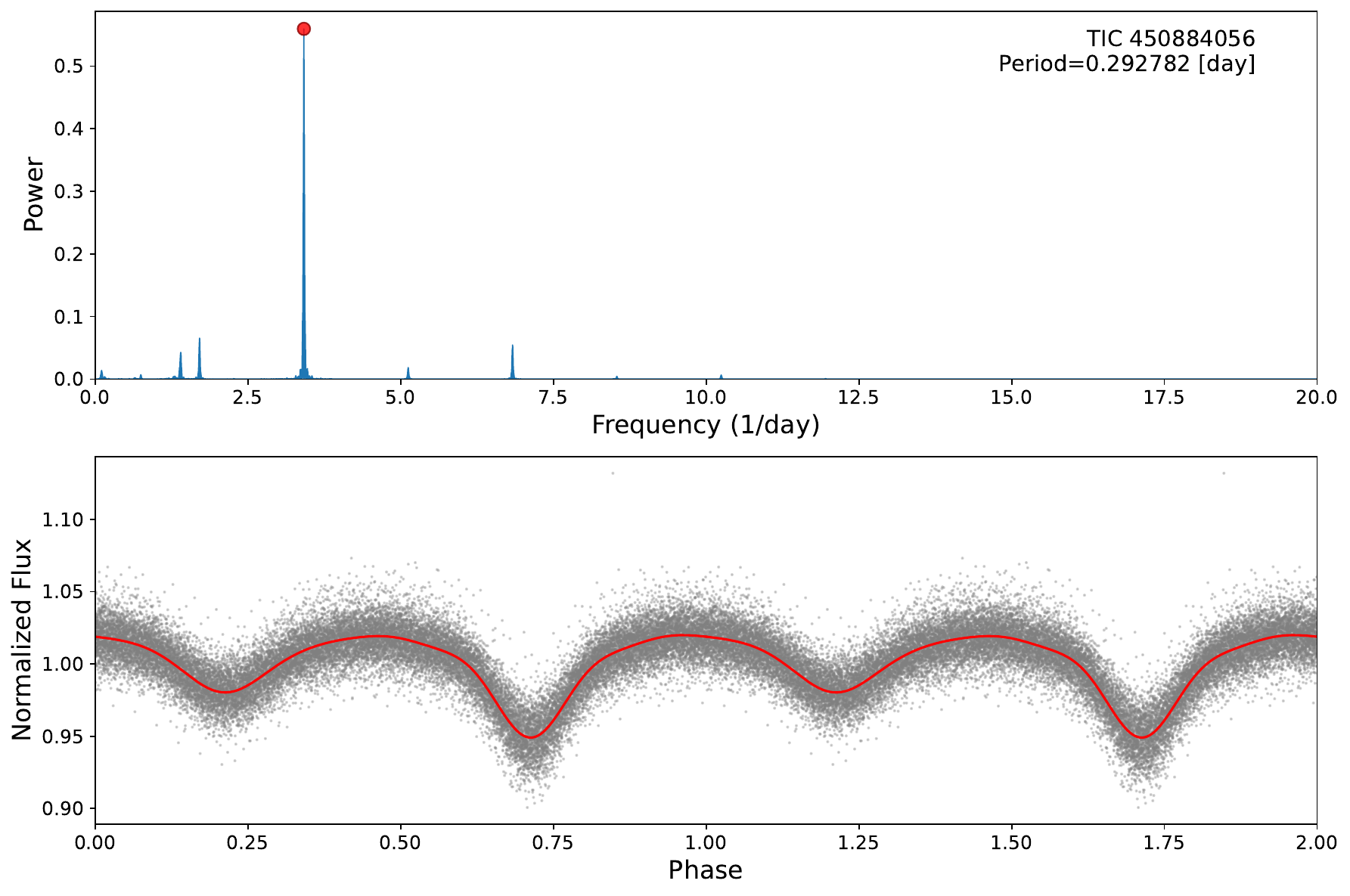}
    \caption{Example TESS light curve of an EB-type eclipsing binary system hosting a BeT star. Top: Lomb-Scargle periodogram (power spectral density, PSD) of the TESS light curve; the red filled circle marks the orbital frequency. Bottom: phased light curve folded on the adopted orbital period; the gray dots show the original data and the red line shows the best-fitting Fourier model.}
    \label{bn}
\end{figure*}

We identified 63 eclipsing binary systems from the TESS light curve morphology analysis.
Their binary classifications and adopted periods are included in the machine-readable catalogue, with representative entries shown in Table \ref{ob}.
Among them, 33 are classified as EA, 17 as EB, and 13 as EW systems.
As an example, Figure \ref{bn} presents an eclipsing binary system hosting a BeT star, which exhibits the characteristic features of an EB-type binary.

These 63 systems should not be interpreted as a measurement of the intrinsic binary fraction of BeT stars.
TESS photometry is primarily sensitive to systems that exhibit eclipses, ellipsoidal variability, or other clear periodic modulations, and it will miss many binaries with unfavourable inclinations or longer orbital periods.
Rather than providing a complete binary census, this analysis demonstrates the utility of the BeT catalogue for identifying TESS-detectable eclipsing binary systems.

\section{DISCUSSION}

\subsection{Sample purity and completeness}\label{purity}

Because our BeT candidates are selected from photometric variability alone, without requiring spectroscopic confirmation, it is essential to assess the purity of the catalogue against other types of variable OBA stars and its completeness against spectroscopically confirmed Be stars. We address these two aspects in turn below.

To evaluate the contamination rate, we first cross-matched our BeT catalogue with the BeSS database \citep{2011AJ....142..149N}, which contains 2,456 spectroscopically confirmed classical Be stars, and found 210 matches. 
The majority of our sample (3,090 sources) are therefore new identifications that were not previously catalogued as Be stars in BeSS. 
For these 3,090 sources, we cross-matched with the ZTF variable-star catalogue of \citet{2020ApJS..249...18C} and found 54 matches, of which two are also identified as eclipsing binaries in our TESS analysis. 
The remaining 52 matches are all classified as semiregular (SR) variables, which are typically evolved red giants or supergiants; these sources occupy the red end of the $W_1-W_2$ colour distribution and can be readily distinguished from the bulk of the BeT sample. We then cross-matched the remaining 3,036 sources with the TESS variable-star catalogue of \citet{2025ApJS..276...57G}, yielding 54 matches, all classified as GCAS ($\gamma$ Cassiopeiae variables). Because GCAS variables are rapidly rotating B-type stars with mass outflow, physically the same phenomenon as classical Be stars, these matches support rather than contaminate the Be nature of our sample.

As a final check, we cross-matched the remaining 2,982 sources with the SIMBAD database. Taking the classifications at face value, the 1,772 sources with usable classification information can be grouped into four categories: 654 (36.9 per cent) have emission-line or Be-type classifications that directly support a Be nature; 156 (8.8 per cent) have classifications that are not in conflict with a Be nature, such as pulsating or eclipsing variables; 112 (6.3 per cent) have classifications that are potentially inconsistent with classical Be stars, such as YSOs, long-period variables, AGB stars, or Wolf--Rayet stars; and the remaining 850 (48.0 per cent) lack informative classifications. 
The SIMBAD cross-match therefore provides an empirical measure of known classification conflicts within the externally classified subset: 112 of the 1,772 sources (6.3 per cent) carry classifications potentially inconsistent with a classical Be interpretation. 
Together with the substantial subset carrying independent Be-related classifications and the spectroscopic and variability-based checks described below, this provides no evidence for a dominant contaminating population in the all-sky Gaia-selected sample.

\begin{figure}
	\includegraphics[width=\linewidth]{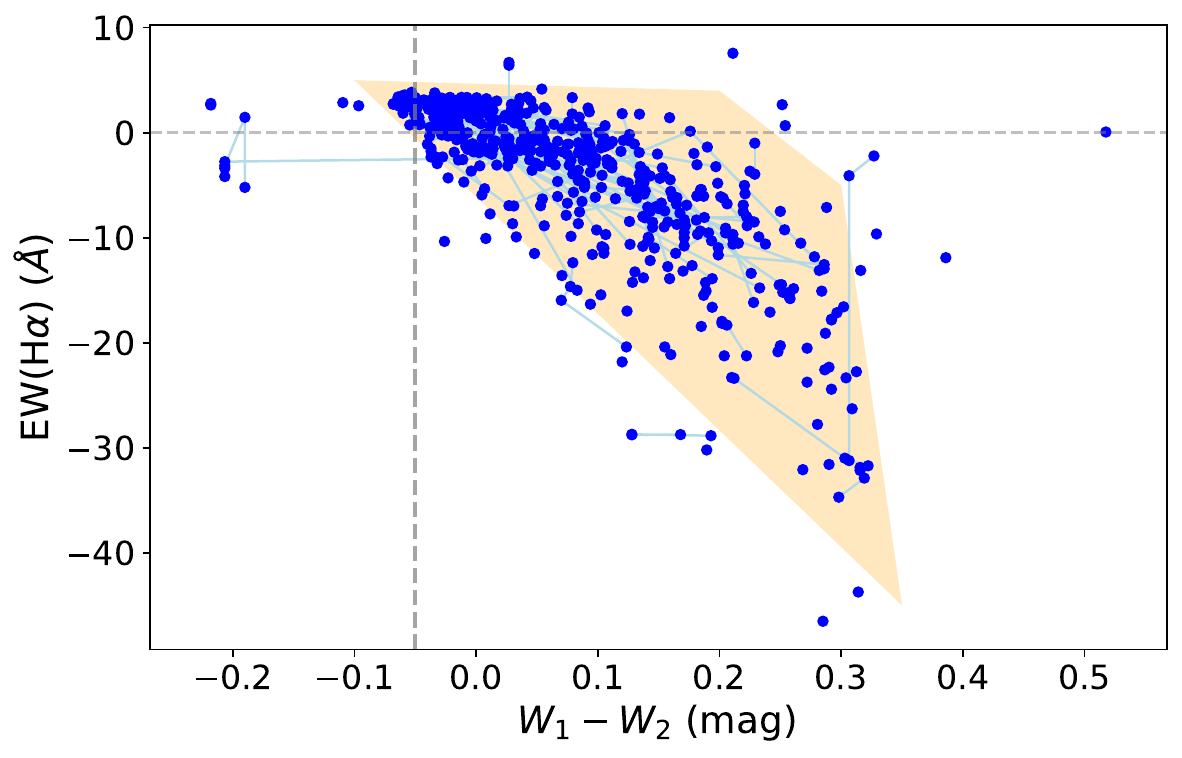}
    \caption{$W_1-W_2$ colour versus H$\alpha$ equivalent width for the BeT stars in this work with LAMOST spectra (blue points). The orange shaded region marks the locus occupied by the 916 spectroscopically confirmed BeT stars of \protect\citet{2024A&A...682A..59J}, and the vertical dashed line indicates the intrinsic colour $(W_1-W_2)_0 \approx -0.05$ mag.}
    \label{color_ew}
\end{figure}

As an additional consistency check, we compared the distribution of our BeT stars in the $W_1-W_2$ colour versus H$\alpha$ equivalent-width plane with that of the 916 spectroscopically confirmed BeT stars of \citet{2024A&A...682A..59J}. 
As shown in Figure \ref{color_ew}, the two samples occupy the same locus, both extending from the intrinsic colour $(W_1-W_2)_0 \approx -0.05$ mag towards redder colours and stronger H$\alpha$ emission, with colours mostly confined to the range 0.0--0.35 mag. 
This agreement supports the interpretation that the majority of our BeT candidates are classical Be stars undergoing disk-related variability, rather than a heterogeneous mixture of other variable OBA types.

\begin{table*}
    \caption{Number of BeT stars in common with \citet{2023ApJS..264...38M}, with the numbers of CBe stars in their study given in parentheses.}
    \centering
    \begin{tabular}{lcccccccc}
    \hline\hline
    \multirow{2}{*}{Sample} & \multicolumn{3}{c}{Secular} & & \multicolumn{3}{c}{Stochastic} & Total \\
    \cline{2-4} \cline{6-8}
    & sin & sin+linear & linear & & burst & drop & irregular \\
    PMS & 2 & 0 & 1 & & 2 & 1 & 2 & 8\\
    Uncertain & 9 & 3 & 6  & & 3 & 4 & 9 & 34\\
    CBe & 11(94) & 15(92) & 11(91) & & 6(18) & 4(36) & 13(128) & 60 (459)\\
    \hline
         &  \\
    \end{tabular}
    \label{mei2023_cross}
\end{table*}

We have further checked whether other types of variable OBA stars could contaminate the BeT sample, including early-type supergiants with time-variable winds, luminous blue variables, and pre-main-sequence (PMS) Herbig Ae/Be stars. 
Following \citet{2024A&A...682A..59J}, we cross-matched our catalogue with the Herbig Ae/Be and classical Be catalogue of \citet{2020A&A...638A..21V} and with the NEOWISE variability study of \citet{2023ApJS..264...38M}. 
The latter cross-match (Table \ref{mei2023_cross}) shows that only 8 sources classified as pre-main-sequence stars are identified as BeT stars, far fewer than the 60 confirmed classical Be stars. 
These cross-matches provide an additional check on contamination from pre-main-sequence and Herbig Ae/Be objects, and support the interpretation that the bulk of our photometrically selected candidates are classical Be stars. 
Possible contamination from other variable-star classes is considered further below.

To assess whether brightening in $W_1$ accompanied by reddening in $W_1-W_2$ and a strong anti-correlation ($r<-0.70$) is a unique identifier of BeT behaviour, we applied our full selection procedure to known non-Be variable stars. 
We extracted all variables classified as non-Be types from the ZTF variable-star catalogue of \citet{2020ApJS..249...18C} and the TESS variable-star catalogue of \citet{2025ApJS..276...57G} (5,874 and 6,802 sources, respectively, after excluding sources already present in our BeT catalogue) and cross-matched them with the WISE light curves. 
In the ZTF sample, 3,875 (66.0 per cent) and 251 (4.3 per cent) sources pass the data-quality and colour-variation steps, respectively; in the TESS sample, the corresponding numbers are 5,130 (75.4 per cent) and 158 (2.3 per cent). 
Crucially, none of the sources passing the colour-variation step in either catalogue satisfies the strong anti-correlation criterion $r<-0.70$. 
This residual-sample test therefore serves as an additional consistency check: it shows that, outside the objects already included in our BeT catalogue, these external variable-star catalogues do not contain an additional population that independently satisfies the full selection criteria.

To estimate the fraction of the classical Be star population that our method captures, we examined how the 2,456 BeSS stars pass through each step of our selection. 
Of these, 1,570 (63 per cent) are present in our Gaia-selected OBA parent sample; the remaining 886 are lost at this initial stage, primarily because they are too faint or lack the required Gaia stellar-parameter estimates. 
Among the sources entering the parent sample, 1,205 (77 per cent) have WISE/NEOWISE light curves that pass our quality cuts, 852 (70 per cent) satisfy the mean magnitude and colour criteria, and 526 (62 per cent) meet the $W_1$ and $W_1-W_2$ amplitude thresholds. 
Applying the correlation criterion ($r<-0.70$) reduces the sample to 210 sources (40 per cent of the amplitude-selected subset), all of which survive visual inspection. 
The overall capture fraction is therefore 210 out of 2,456, or 8.5 per cent of the BeSS Be stars, while the detectable BeT fraction among the WISE-eligible known Be stars is 210 out of 852, or 24.6 per cent.

This relatively low recovery rate is a direct consequence of our selection function, which is designed to identify \textit{transition} events, namely stars that exhibit clear, correlated variations in infrared brightness and colour on timescales longer than a few months. 
The make-up of our sample is governed primarily by the WISE/NEOWISE observing strategy (cadence, baseline and depth). 
Our method inevitably misses Be stars with stable disks that show little long-term infrared variability, Be stars whose outbursts are shorter than, or poorly sampled by, the $\sim$ 6-month WISE cadence, and Be stars that remained in a prolonged quiescent state throughout the WISE/NEOWISE baseline. 
These sources are genuine classical Be stars, but they do not exhibit the photometric signatures required for identification as BeT stars. 
Conversely, the recovered sources constitute a uniformly selected sample of Be stars that are actively undergoing disk formation or dissipation. 
We emphasise that the scientific value of the BeT catalogue lies in this dynamic information: whereas a conventional Be classification records the presence of a disk at one or a few spectroscopic epochs, the BeT sample provides direct evidence of disk state changes, supplying the observational basis for studying disk viscosity, mass-injection history, and the duty cycle of Be outbursts. 
The duty-cycle and timescale information in this catalogue will also be useful for planning and interpreting future time-domain photometric surveys: for example, the Tianyu survey \citep{2024arXiv2404.07149F} and the Nancy Grace Roman Space Telescope \citep{2015arXiv150303757S} will provide dense optical light curves over several years, with which Be-star disk activity can be monitored and the statistics derived here can be tested.

Together, the external classification checks, spectroscopic validation, and the characterised selection function support the use of the catalogue for statistical studies of Be-disk evolution.
The newly identified BeT stars substantially expand the sample of actively transitioning Be stars and extend its distance coverage relative to \citet{2024A&A...682A..59J}, as shown in Figure \ref{dis}.

\subsection{RV Variable Sources}

\begin{figure*}
	\centering
	\includegraphics[width=\linewidth]{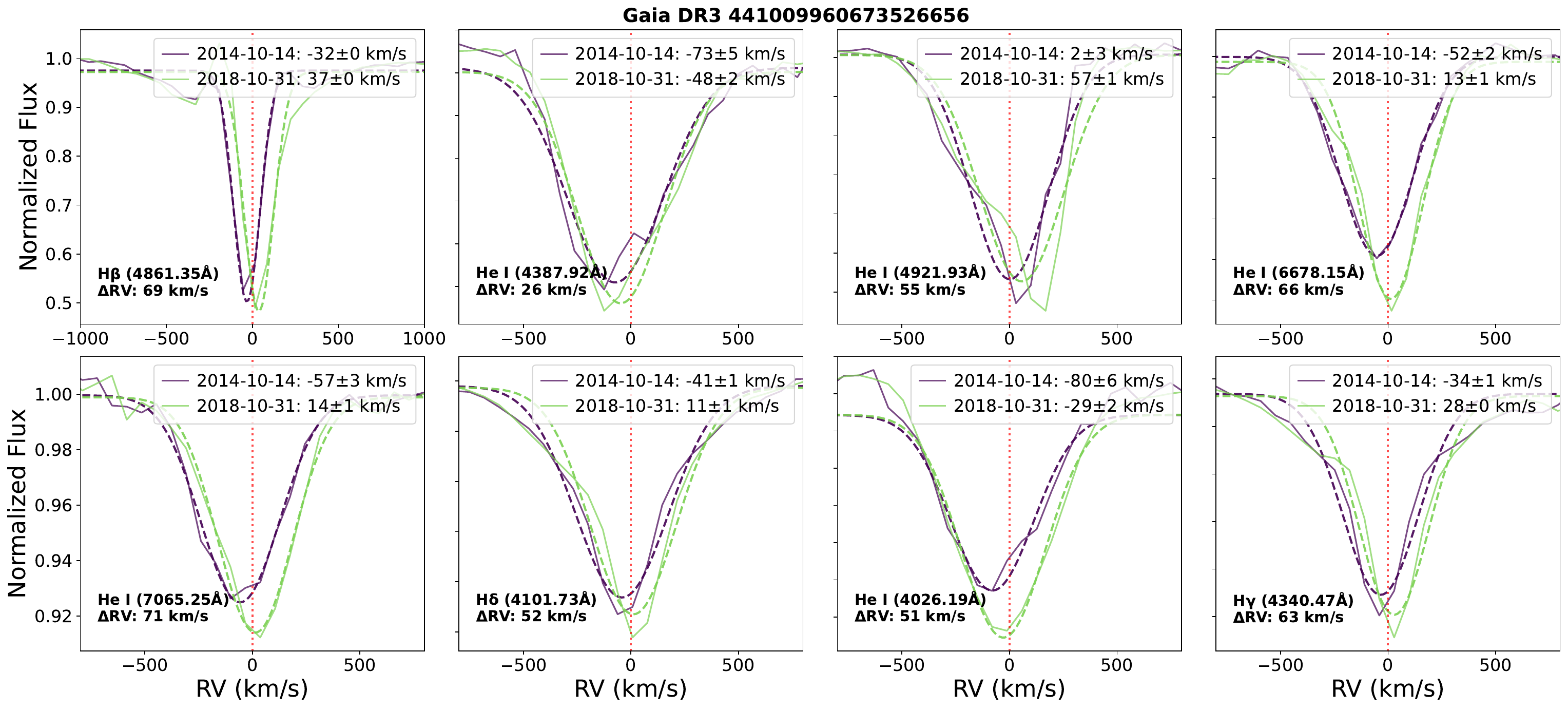}
    \caption{Normalized flux as a function of RV for multiple spectral lines of LAMOST J030246.39+523958.8 (Gaia DR3 441009960673526656). The panels show H lines (4861.35, 4101.73, 4340.47\,\AA) and several He\,I lines (4026.19, 4387.92, 4921.93, 6678.15, 7065.25}\,\AA). Each panel corresponds to one spectral line, labelled with its rest wavelength. Vertical dashed lines indicate the zero RV. For each line, we report the RV difference between the two epochs and the RV measured at each epoch with its $1\sigma$ uncertainty. All flux profiles have been normalized to the local continuum.
    \label{rv1}
\end{figure*}

\begin{figure*}
	\centering
	\includegraphics[width=\linewidth]{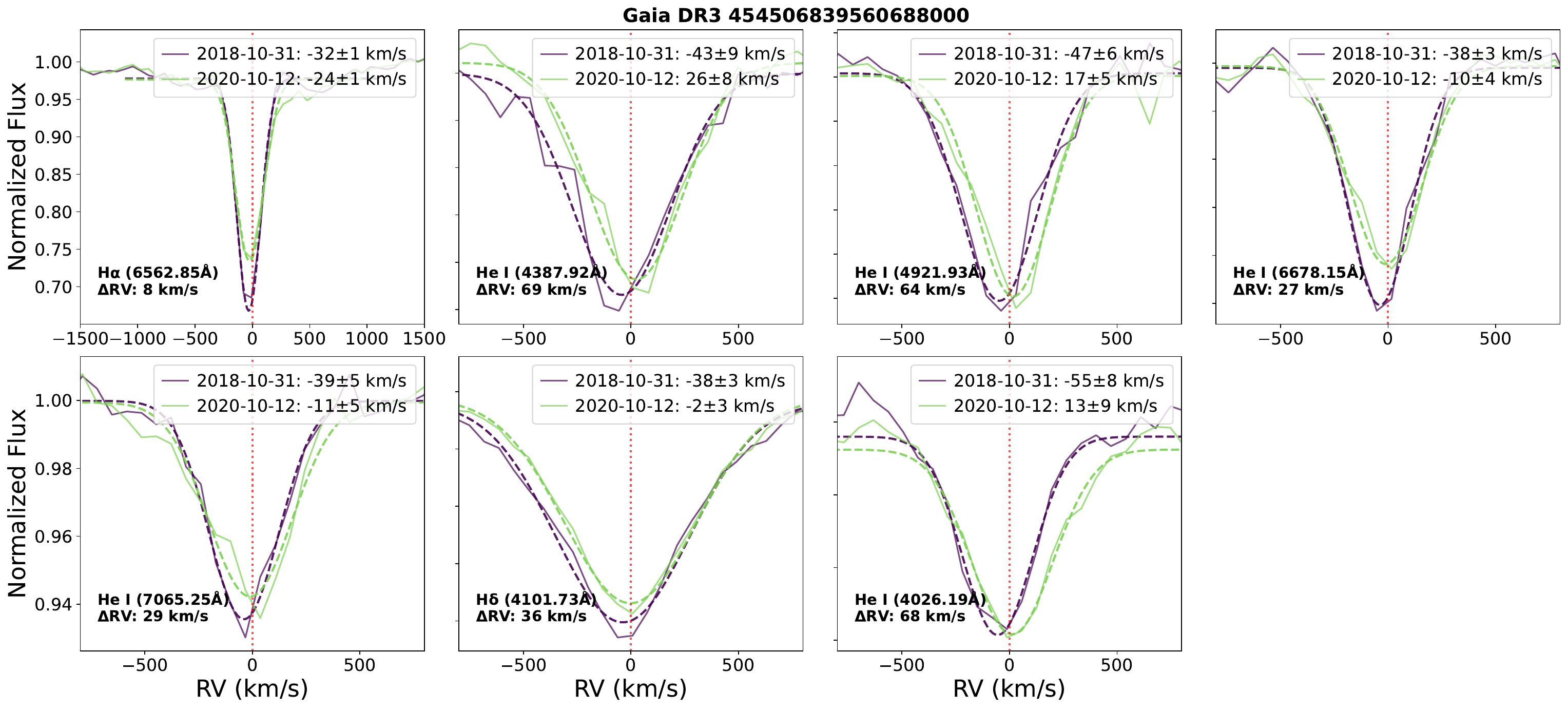}
    \caption{Same as Fig. \ref{rv1}, but for LAMOST J025035.09+560447.5 (Gaia DR3 454506839560688000), showing H lines (6562.85, 4101.73 \AA) and He\,I lines (4387.92, 4026.19, 4921.93, 6678.15, 7065.25\,\AA).}
    \label{rv2}
\end{figure*}

For the remaining 87 sources with multiple LRS observations, we conducted RV analysis. 
Only two sources met the criterion of an epoch-to-epoch RV shift exceeding 20 km s $^{-1}$ : LAMOST J030246.39+523958.8 and LAMOST J025035.09+560447.5.
Their line-by-line RV measurements are shown in Figures \ref{rv1} and \ref{rv2}.
In LAMOST J025035.09+560447.5, the H$\alpha$ profile does not show a clear centroid shift, whereas H$\gamma$ (4340.47\,\AA), H$\beta$ (4861.35\,\AA), H$\delta$ (4101.73\,\AA), and several He\,I lines show measurable epoch-to-epoch centroid shifts. 
The H$\alpha$ profile is affected by disk emission and therefore does not provide a reliable stellar RV in this case.
We therefore classify the object as an RV variable based on the shifts measured from the other, less emission-contaminated photospheric lines.

To test whether these RV shifts are likely to originate from binary motion, we inspected the available TESS light curves of the two RV variable sources. 
Neither source exhibits clear photometric signatures of binarity, such as eclipses or ellipsoidal modulation. 
For comparison, we also examined the TESS data of the 13 binary candidates reported by \cite{2026ApJS..283...59A}; most of them show clear binary-like light curve morphologies, whereas one candidate, J035933.84+555751.1, similarly shows no such TESS signature.
Thus, the absence of TESS binary signatures does not rule out binarity, but it weakens the interpretation that the detected RV shifts are unambiguous orbital signals.

\begin{figure*}
	\centering
	\includegraphics[width=\linewidth]{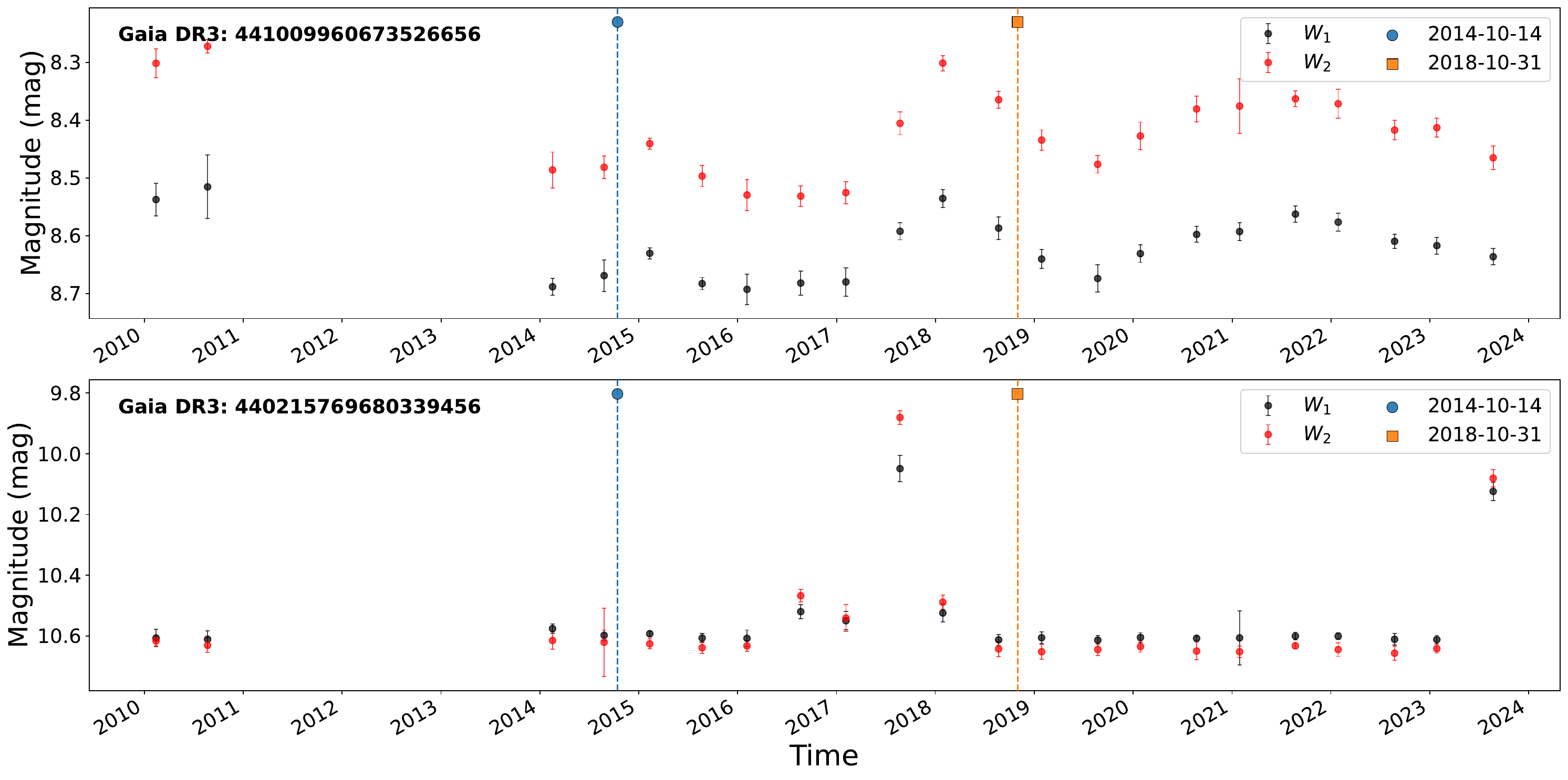}
\caption{WISE mid-infrared light curves for one RV variable source and one comparison source: LAMOST J030246.39+523958.8 (top panel, Gaia DR3 441009960673526656) and LAMOST J030551.37+522900.5 (bottom panel, Gaia DR3 440215769680339456). Vertical dashed lines mark the epochs of LAMOST LRS observations.}
\label{rv12}
\end{figure*}

\begin{figure*}
	\centering
	\includegraphics[width=\linewidth]{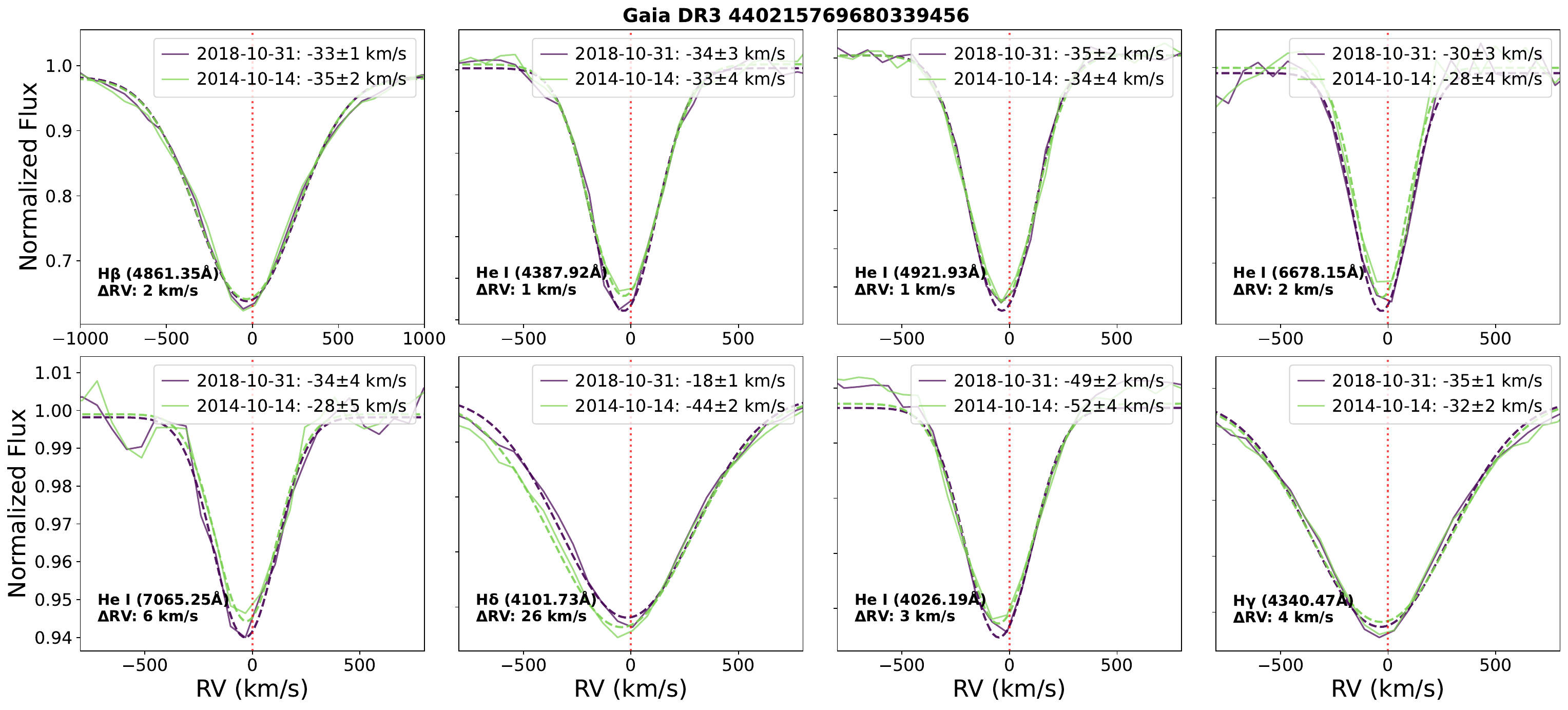}
    \caption{Same as Fig. \ref{rv1}, but for LAMOST J030551.37+522900.5 (Gaia DR3 440215769680339456) showing H lines (4861.35, 4101.73, 4340.47 \AA) and He\,I lines (4026.19, 4387.92, 4921.93, 6678.15, 7065.25\,\AA).}
    \label{rv3}
\end{figure*}

For LAMOST J030246.39+523958.8 (Figure \ref{rv1}), the available LAMOST epochs fall during an active or outbursting state in its WISE light curve (Figure \ref{rv12}).
This timing suggests that outburst-related line profile variability may contribute to the measured RV shifts.
In contrast, the comparison source LAMOST J030551.37+522900.5 (shown in Figure \ref{rv3}) shows no significant RV shifts at epochs that appear to correspond to quiescent or disk dissipation phases in WISE. 
Although this comparison is only illustrative, it indicates that RV variability in active Be stars is not always an unambiguous indicator of binarity.

This interpretation is further supported by the line-dependent nature of the measured RV shifts.
If the shifts were purely caused by orbital motion, the clean photospheric absorption lines would be expected to show broadly consistent velocity changes between epochs. 
Instead, some lines show clearer shifts than others, suggesting that line-profile variability associated with disk activity or outburst-related changes may affect the measured centroids. 
Similar behaviour has been reported in earlier studies of classical Be stars.

A relevant example is HD 49330. 
\citet{2009A&A...506..103F} reported epoch-dependent RV shifts and line-profile variability in this Be star, and argued that the observed RV variations were unlikely to be caused by orbital motion because a global shift of all lines would then be expected. 
Together with the CoRoT photometric analysis of \citet{2009A&A...506...95H} and the pulsation modelling of \citet{2020A&A...644A...9N}, this case illustrates how pulsations, outflow episodes, and disk-related line-profile changes can complicate the interpretation of RV variability in active Be stars. 
We therefore conclude that although RV monitoring remains a powerful method for detecting binaries, RV variability in actively outbursting Be star can also trace disk-related line profile variability.

\section{Conclusions}
\label{conclusions}
We have constructed an all-sky catalogue of Be transition stars by combining Gaia DR3 astrometry and photometry with multi-epoch mid-infrared photometry from WISE and NEOWISE-R. 
Starting from a Gaia-selected parent sample of candidate OBA stars, we identified BeT candidates through correlated variability in $W_1$ magnitude and $W_1-W_2$ colour, which traces the mid-infrared brightening and reddening associated with disk formation, and the reverse trend during disk dissipation.
The final catalogue comprises 3,300 BeT stars, including 2,743 newly identified objects, and provides a large sample for studying the variability properties and spatial distribution of Be outburst activity.

We measured the durations of outburst and decay phases from the WISE light curves, finding median values of 459 and 478 days, respectively. 
These timescales are broadly consistent with previous WISE-based measurements and are compatible with the long disk-evolution timescales expected in the viscous decretion disk model,
while the wide duration distribution reflects the diversity of disk evolution among Be stars.
As initial applications of the catalogue, we used TESS light curves to identify 63 TESS-detectable eclipsing binary systems hosting Be stars, classified into 33 EA, 17 EB, and 13 EW systems. 
We also used multi-epoch LAMOST spectra to examine H$\alpha$ emission and candidate RV variability.
The LAMOST analysis suggests that RV shifts in active Be stars are not necessarily caused by binary motion, because disk-related line profile variability during outburst phases can also affect measured line centroids.

Overall, our results demonstrate that correlated mid-infrared magnitude and colour variability provides a useful diagnostic of circumstellar disk evolution, and the publicly released catalogue will support future statistical studies of disk formation, mass loss, binarity, pulsation, and outbursts in classical Be stars.

\section*{Acknowledgements}

This work was supported by the National Natural Science Foundation of China (NSFC) through grants 12233013, 12503045, 12573023, 12573035. 
G.L. acknowledges the support of the Australian Research Council through the DECRA project DE250100773. 
We used data from the European Space Agency (ESA) mission Gaia (\url{https://www.cosmos.esa.int/gaia}), processed by the Gaia Data Processing and Analysis Consortium (DPAC, \url{ https://www. cosmos.esa.int/web/gaia/dpac/consortium}). 
Funding for the DPAC has been provided by national institutions, in particular the institutions participating in the Gaia Multilateral Agreement. 
This publication makes use of data products from ALLWISE (\url{https://wise.ssl.berkeley.edu/index.html}) and NEOWISE (\url{https://neowise.ipac.caltech.edu/}), which are projects of the Jet Propulsion Laboratory/California Institute of Technology. 
ALL-WISE and NEOWISE are funded by the National Aeronautics and Space Administration. 
This work has made use of LAMOST spectra. 
Guoshoujing Telescope (LAMOST; \url{http://www.lamost.org/public/}) is a National Major Scientific Project built by the Chinese Academy of Sciences (CAS). 
Funding for the project has been provided by the National Development and Reform Commission. 
LAMOST is operated and managed by the National Astronomical Observatories, CAS.

\section*{Data Availability}
The BeT catalogue and associated WISE light curves are publicly available via China VO at \texttt{https://nadc.china-vo.org/res/r101801/.}
The Gaia, WISE/NEOWISE, TESS, and LAMOST data used in this work are available from their respective public archives.
All additional derived data products will be shared on reasonable request to the corresponding author.

\bibliographystyle{mnras}
\bibliography{example}

\appendix
\section{ADQL Query for OBA Star Selection}
\label{adql}

The following ADQL query was used to retrieve candidate O-, B-, and A-type stars from the Gaia DR3 archive, based on effective temperature and spectral type criteria, while applying an absolute magnitude and colour cut to remove nearby cool dwarf contaminants. 
The query combines the \texttt{gaiadr3.gaia\_source} table with the \texttt{gaiadr3.astrophysical\_parameters} table to obtain astrometric quantities, photometry, and Gaia DR3 stellar parameter estimates.
Spatial constraints on right ascension and declination were applied to enable segmented all-sky retrieval; the example below shows one such sky segment.

\begin{lstlisting}[style=sqlstyle]
SELECT g.*, ap.*
FROM gaiadr3.gaia_source AS g
INNER JOIN gaiadr3.astrophysical_parameters AS ap 
ON g.source_id = ap.source_id 
WHERE 
((ap.teff_gspphot > 7000 
     AND ap.spectraltype_esphs IN ('O', 'B', 'A') 
     AND ap.teff_esphs IS NULL)
    OR (ap.teff_esphs > 10000 AND ap.teff_gspphot > 8000)
    OR (ap.teff_esphs > 10000 
        AND ap.teff_esphs < 50000 
        AND ap.teff_gspphot IS NULL))
AND 
    POWER(g.parallax / 100.0, 2) < POWER(10.0, 2.0 - g.phot_g_mean_mag + 1.8 * g.bp_rp)
AND g.ra BETWEEN 0 AND 10  
AND g.dec BETWEEN 0 AND 10
\end{lstlisting}

\bsp	
\label{lastpage}
\end{document}